\documentclass[12pt]{article}
\usepackage{graphicx}
\def \d {{\rm d}}

\def \k {k}

\newcommand{\imgtype}{s}  

\newcommand{\Ua}{{U_{\mathrm{a}}}}
\newcommand{\Va}{{V_{\mathrm{a}}}}
\newcommand{\Ta}{{T_{\mathrm{a}}}}
\newcommand{\Ra}{{R_{\mathrm{a}}}}
\newcommand{\Ha}{{{\cal H}_{\mathrm{a}}}} 

\newcommand{\Uo}{{U_{\mathrm{o}}}}
\newcommand{\Vo}{{V_{\mathrm{o}}}}
\newcommand{\To}{{T_{\mathrm{o}}}}
\newcommand{\Ro}{{R_{\mathrm{o}}}}
\newcommand{\Ho}{{{\cal H}_{\mathrm{o}}}} 

\newcommand{\zBP}{{\hat\zeta}}
\newcommand{\rBP}{{\hat\rho}}
\newcommand{\TBP}{{\hat T}}
\newcommand{\ZBP}{{\hat Z}}
\newcommand{\RBP}{{\hat R}}

\newcommand{\zBR}{\zeta}
\newcommand{\TBR}{T}
\newcommand{\ZBR}{Z}

\newcommand{\tE}{{\tilde t}}
\newcommand{\uE}{{\tilde u}}
\newcommand{\vE}{{\tilde v}}

\newcommand{\ph}{\varphi}
\newcommand{\tht}{\vartheta}

\newcommand{\scri}{{\mathcal{I}}}

\newcommand{\sign}{\mathop{\mathrm{sign}}}

\begin{document}

\title{\bf Interpreting the $C$-metric}

\author{J. B. Griffiths$^1$, P. Krtou\v{s}$^2$ and J. Podolsk\'y$^2$
\\ \\ \small
$^1$Department of Mathematical Sciences, Loughborough University, \\
\small Loughborough,  Leics. LE11 3TU, U.K. \\ 
\small $^2$Institute of Theoretical Physics, Charles University in Prague,\\
\small V Hole\v{s}ovi\v{c}k\'ach 2, 18000 Prague 8, Czech Republic.
\\ \\
\small E-mail: J.B.Griffiths@lboro.ac.uk, Pavel.Krtous@mff.cuni.cz \\ 
\small and podolsky@mbox.troja.mff.cuni.cz}

\date{\today}
\maketitle

\begin{abstract}
\noindent
The basic properties of the $C$-metric are well known. It describes a pair of causally separated black holes which accelerate in opposite directions under the action of forces represented by conical singularities. However, these properties can be demonstrated much more transparently by making use of recently developed coordinate systems for which the metric functions have a simple factor structure. These enable us to obtain explicit Kruskal--Szekeres-type extensions through the horizons and construct two-dimensional conformal Penrose diagrams. We then combine these into a three-dimensional picture which illustrates the global causal structure of the space-time outside the black hole horizons. Using both the weak field limit and some invariant quantities, we give a direct physical interpretation of the parameters which appear in the new form of the metric. For completeness, relations to other familiar coordinate systems are also discussed.

\end{abstract}

\section{Introduction}

The purpose of this paper is to clarify certain aspects of the physical interpretation of the vacuum space-time that was referred to as the $C$-metric in the classic review of Ehlers and Kundt~\cite{EhlersKundt62} -- a label that has been widely used ever since. In fact, the static form of this solution was originally found by Weyl in 1917~\cite{Weyl17} and subsequently rediscovered many times. Its basic properties were first interpreted by Kinnersley and Walker~\cite{KinWal70} (see also Bonnor~\cite{Bonnor83}), who showed that its analytic extension represents a pair of black holes which accelerate away from each  other due to the presence of strings or struts that are represented by conical singularities. The radiative properties of this space-time were investigated by Farhoosh and Zimmerman~\cite{FarZim80b} and by Bi\v{c}\'ak~\cite{Bicak85}, and its asymptotic properties by Ashtekar and Dray~\cite{AshDra81}. For reviews of more recent work see e.g. \cite{Pravdas00} and \cite{LetOli01}.

Although the line element for the $C$-metric has been expressed using many different coordinate systems, it is appropriate to initially quote it in the well known form 
 \begin{equation} 
 \d s^2={1\over A^2({\rm x}+{\rm y})^2}\left( -{\rm F}\,\d{\rm t}^2 +{\d{\rm y}^2\over{\rm F}}
  +{\d{\rm x}^2\over{\rm G}} +{\rm G}\,\d\phi^2  \right) ,
 \label{PDCmetric} 
 \end{equation}   
 where ${\rm G}$ and ${\rm F}$ are cubic functions of ${\rm x}$ and ${\rm y}$ respectively:
 \begin{equation} 
 {\rm G}=1-{\rm x}^2-2MA\,{\rm x}^3, \qquad {\rm F}=-1+{\rm y}^2-2MA\,{\rm y}^3. 
 \label{PDPQ} 
 \end{equation} 
 This solution contains two constant parameters $M$ and $A$. It reduces to a form of Minkowski space when $M=0$, but has no convenient limit as $A\to0$. It will be shown below that it is only after certain rescalings that these parameters acquire specific physical interpretations.

It may be observed that the functions ${\rm G}({\rm x})$ and ${\rm F}({\rm y})$ are related by the condition that ${\rm F}(w)=-{\rm G}(-w)$. It follows that these functions must have the same structural properties such as the same number of (related) roots. In particular, ${\rm G}$ and ${\rm F}$ will each possess three distinct real roots if $27M^2A^2<1$, and we will assume that this condition is satisfied. It may also be noted that, in the form (\ref{PDCmetric}) with (\ref{PDPQ}), the coordinate freedom to make a linear transformation in ${\rm x}$ and ${\rm y}$ has been used to remove the linear terms in ${\rm G}$ and ${\rm F}$.

Although the metric (\ref{PDCmetric}) has been extensively used in the literature, a significant simplification was recently introduced by Hong and Teo \cite{HongTeo03}. Their innovation was to use the freedom to apply a linear transformation to the coordinates ${\rm x}$ and ${\rm y}$, not to remove the linear components in the cubic functions as in (\ref{PDPQ}) but, rather, to use the transformation (together with a rescaling of ${\rm t}$ and $\phi$ and the parameters $M$ and $A$) in such a way that the root structure of the cubics is expressed in a very simple form. Specifically, assuming the existence of three roots with the coefficient of the cubic term being negative, transformations are applied to fix the roots of the metric function in $x$ at $+1$, $-1$ and $-1/(2\alpha m)$. In order to preserve the order of the roots, it is necessary to assume that $0<2\alpha m<1$. In this way, Hong and Teo have re-expressed the line element (\ref{PDCmetric}) in the form  
 \begin{equation} 
 \d s^2={1\over\alpha^2(x+y)^2}\left( -F\,\d \tau^2
  +{\d y^2\over F} +{\d x^2\over G} +G\,\d\varphi^2  \right) ,
 \label{HTCmetric} 
 \end{equation}  
 where 
 \begin{equation} 
 G=(1-x^2)(1+2\alpha mx), \qquad F=-(1-y^2)(1-2\alpha my). 
 \label{HTPQ} 
 \end{equation}

To maintain a Lorentzian signature of the metric (\ref{HTCmetric}), it is necessary that $G>0$, which implies that the coordinate $x$ must be constrained to lie between appropriate roots of the cubic function $G$. On the other hand, there is no constraint on the sign of $F$ (static regions occur when $F>0$). It can also be seen from (\ref{HTCmetric}) that $x+y=0$ corresponds to conformal infinity. Thus a physical space-time must satisfy either $x+y>0$ or $x+y<0$. With these constraints on the coordinates ${x,y}$, the metric (\ref{HTCmetric}) can still describe four qualitatively different space-times depending on particular choices of the coordinate ranges. We will here restrict attention to the physically most important case ${x\in(-1,1)}$ and ${x+y>0}$, in which the metric describes the space-time of two accelerated black holes.

In the form of the metric~(\ref{HTCmetric}), the roots of $G$ and $F$ have very simple explicit expressions, while still satisfying the condition that $F(w)=-G(-w)$. In fact, their simplified root structure is of considerable assistance when it comes to performing calculations, and also helps in the physical interpretation of this solution. However, particularly for convenience near the black holes, closely related spherical-like coordinates will be introduced in section~\ref{sc:sphcoords}. In either of these forms, the simple factor structure of the metric functions enables Kruskal--Szekeres-type transformations to be performed. These are given explicitly for the first time in section~\ref{sc:horizons} and clarify the extensions through the horizons. They also enable conformal Penrose diagrams to be constructed. After first discussing the weak field limit in sections~\ref{sc:WeakField} and~\ref{sc:glstrWFL}, these two-dimensional diagrams are combined in section~\ref{sc:glstr} to form a three-dimensional picture which nicely represents the global causal structure of the regions outside the black hole horizons. Some particular invariant physical and geometrical quantities are reviewed in section~\ref{sc:PhysProp}. Finally, explicit relations to other familiar coordinate systems are presented in sections~\ref{sc:BoostRot} and~\ref{sc:tphconst}, further demonstrating the advantages of the simplified root structure of the new metric functions.

\section{The $C$-metric in spherical-type coordinates} 
\label{sc:sphcoords}
    Although the $x,y$ coordinates of (\ref{HTCmetric}) and (\ref{HTPQ}) are useful when performing calculations and in analysing the global structure the space-time, their physical interpretation can be made more clear if we introduce closely related coordinates ${r,\,\theta}$ which will play a role of spherical coordinates around the black holes. Since we restricted $x$ to the range $-1<x<1$, it is natural to introduce the coordinate transformation 
 \begin{equation} 
 x=\cos\theta, \qquad y={1\over\alpha r}, \qquad \tau=\alpha\,t ,
 \label{xytothetar} 
 \end{equation} 
 where $\theta\in(0,\pi)$. With (\ref{xytothetar}), the metric becomes 
 \begin{equation} 
 \d s^2={1\over(1+\alpha r\cos\theta)^2} \left( -Q\,\d t^2 
 +{\d r^2\over Q} +{r^2\,\d\theta^2\over P} +P\,r^2\sin^2\theta\,\d\varphi^2 \right), 
 \label{BLCmetric} 
 \end{equation}  
 as in the nonrotating and uncharged case of that described in \cite{GriPod05}, where 
 \begin{equation} 
  P=1+2\alpha m\cos\theta, \qquad 
  Q=(1-\alpha^2r^2)\Big(1-{2m\over r}\Big). 
 \label{BLPQ} 
 \end{equation} 
It can clearly be seen that this represents a family of solutions with two parameters $m$ and $\alpha$ which (in contrast to the metric (\ref{HTCmetric})) has the important property that it reduces precisely to the familiar form of the spherically symmetric Schwarzschild solution when ${\alpha=0}$.

Relative to a null tetrad that is naturally adapted to (\ref{BLCmetric}), it can be shown that the only nonzero component of the curvature tensor is that given by 
 \begin{equation} 
 \Psi_2 =-m\left({1\over r}+\alpha\cos\theta\right)^3 =-m\alpha^3(x+y)^3. 
 \label{CWeyl} 
 \end{equation} 
 This implies that the space-time is of algebraic type~D and that (except for the case when $m=0$ which is flat) a {\em curvature singularity} occurs at $r=0$. Notice also that the space-time becomes flat for ${x+y\to0}$, i.e., near conformal infinity. The roots of $Q$ at $r=2m$ and $r=1/\alpha$ are coordinate singularities which correspond to Killing horizons.

Since this solution is a one-parameter generalization of the Schwarzschild space-time, it is natural to continue to interpret the constant $m$ (which we assume to be positive) as the parameter characterizing the mass of the source, and $r>0$ as a Schwarzschild-like radial coordinate with a {\em black hole horizon} $\Ho$ at $r=2m$. 
The remaining question therefore concerns the character of the new parameter~$\alpha$ (which must also now be positive) and the horizon~$\Ha$ at $r=1/\alpha$. We will argue that this solution describes an accelerating black hole in which $\alpha$ can be interpreted as the acceleration. Several reasons for this interpretation will be given. The first is that the horizon at $r=1/\alpha$ has the character of an {\em acceleration horizon} associated with a boost symmetry. It will also be shown in the section~\ref{sc:WeakField} that $\alpha$ is exactly the acceleration of the corresponding source in the weak field limit.

\begin{figure}[ht]     
\centerline{
\includegraphics[scale=1.0]{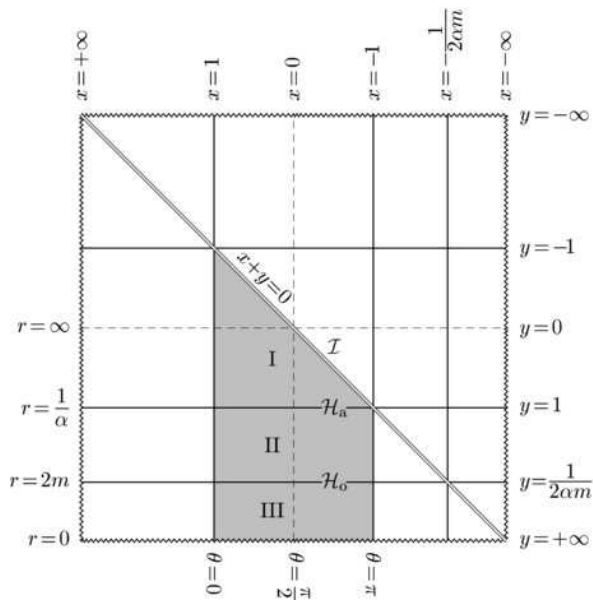}}
\vspace*{-1ex}
\caption{\small 
The full ranges of the $x,y$-coordinates are indicated together with the roots of the cubic functions $G(x)$ and $F(y)$ given by (\ref{HTPQ}). The diagonal $x+y=0$ corresponds to conformal infinity~$\scri$. The shaded region represents the part of the space-time that has most physical significance and is discussed here. In the $r,\theta$-coordinates, $\theta=0$ and $\theta=\pi$ correspond to coordinate poles. For $\theta=$~const., $r$ increases from zero at the singularity through the black hole horizon~$\Ho$ and the acceleration horizon~$\Ha$ and then either to conformal infinity or to the coordinate limit $r=\infty$, through which the space-time can be extended. }
\label{PDcoords}
\end{figure}

The relation between the $r,\theta$ and $x,y$ coordinates, the roots of $Q$, and the constraint representing conformal infinity are illustrated in figure~\ref{PDcoords}. 
As indicated, the horizons divide the space-time into three distinct domains labelled I, II, and III. In the domain II, the Killing vector ${\partial_t}$ is timelike and the space-time static. The coordinate $t$ is temporal here, while ${r}$ is spatial. In domains I and III, the character of the coordinates ${t}$ and ${r}$ is interchanged.

The ${C}$-metric space-time also admits a second Killing vector ${\partial_\ph}$ which is always spatial and corresponds to the rotational symmetry. It can be seen from the metric (\ref{BLCmetric}) that the roots of the metric coefficient ${g_{\ph\ph}}$, where the norm of the Killing vector ${\partial_{\ph}}$ vanishes, correspond to poles $\theta=0$ and $\theta=\pi$ of the coordinate ${\theta}$. These identify the {\em axis of symmetry} of the space-time. The coordinate $\ph$ is thus taken to be periodic with the range ${\ph\in(-\pi C,\pi C)}$, where the precise value of the positive parameter ${C}$ will be discussed below in section \ref{sc:PhysProp}.

As already mentioned, ${x+y=0}$ corresponds to conformal infinity and we assume that ${y>-x}$ (see the shaded region illustrated in figure~\ref{PDcoords}). However, in our spherical-type coordinates, the region near conformal infinity is problematic. For \hbox{${\pi\over2}<\theta<\pi$}, conformal infinity is simply given by ${r=-(\alpha\cos\theta)^{-1}}$, and we take \hbox{$0<r<(\alpha|\cos\theta|)^{-1}$}. For ${0<\theta<{\pi\over2}}$, conformal infinity is not reached even for $r\to\infty$. To fully cover this asymptotic region of the space-time, it is better to revert to the $x,y$-coordinates.

\section{Extensions across the horizons}
\label{sc:horizons}
The irregularity of the coordinate ${r}$ on the horizons can be seen from the fact that the metric (\ref{BLCmetric}) is singular at $r=1/\alpha$ and at $r=2m$. However, these are merely coordinate singularities and can be removed by suitable coordinate transformations. It turns out that the complete space-time manifold consists of (infinitely) many distinct domains of types I, II, or III which are connected at the horizons. The continuity of the manifold across each horizon can be explicitly demonstrated by introducing suitable coordinates.

This can be achieved by first introducing the double null coordinates
\begin{equation}\label{uvdef}
 u=-r_\star-t, \qquad v=-r_\star+t,
\end{equation} 
 keeping ${\theta}$ and ${\ph}$ unchanged, where ${r_\star}$ is a tortoise coordinate satisfying ${\d r_\star=Q^{-1}\d r}$, i.e.,
 \begin{equation} 
 \label{tortr}
\alpha r_\star  = \k_{\mathrm{c}}\log|1+\alpha r| 
  +\k_{\mathrm{a}}\log|1-\alpha r| 
  +\k_{\mathrm{o}}\log\left|1-\frac r{2m}\right|\;,
 \end{equation} 
 where 
 \begin{equation} 
 \k_{\mathrm{c}}={\frac1{2(1+2\alpha m)}}\;,\quad
 \k_{\mathrm{a}}={-\frac1{2(1-2\alpha m)}}\;,\quad
 \k_{\mathrm{o}}={\frac{2\alpha m}{1-4\alpha^2m^2}}\;.
\end{equation}
 The null coordinates $u,v$ take infinite values at each horizon. To analyse the behaviour near each horizon, we therefore make the rescaling 
\begin{equation}\label{UVuvtransf}
  U=(-1)^i(-\sign\k)\,\exp\Bigl(-\frac{\alpha u}{2\k}\Bigr)\;,\quad
  V=(-1)^j(-\sign\k)\,\exp\Bigl(-\frac{\alpha v}{2\k}\Bigr)\;,
\end{equation} 
 where $i$ and $j$ are appropriately chosen integers and ${\k=\k_{\mathrm{a}}}$ or ${\k=\k_{\mathrm{o}}}$ near the acceleration or black hole horizons respectively. We can then introduce Kruskal--Szekeres-type coordinates ${T,\,R}$ at each horizon where 
\begin{equation}\label{KruskCoor}
T= {\textstyle\frac12}(V+U)\;,\quad R={\textstyle\frac12}(V-U)\;.
\end{equation}

First, let us consider the space-time {\em near the acceleration horizon}~$\Ha$ at $r=1/\alpha$ and put ${\k=\k_{\mathrm{a}}}$. We start with the asymptotic time-dependent domain I with integers ${(i,j)=(0,0)}$, and add the subscript ``${\scriptstyle\mathrm{a}}$'' to the coordinates ${U,V}$ and ${T,R}$. Here ${\Ua}$ and ${\Va}$ take positive values with ${\Ua=0}$ or ${\Va=0}$ at the acceleration horizon. We can then extend the coordinates across the horizon to {\em two} static regions with negative values of ${\Ua}$ or ${\Va}$. These correspond to two domains of type II with ${(i,j)=(0,-1)}$ and ${(i,j)=(-1,0)}$, which can subsequently be extended to another domain of type I with ${\Ua}$ and ${\Va}$ both negative and ${(i,j)=(-1,-1)}$. In the domains of type~I, the Kruskal--Szekeres-type coordinates are explicitly given by
\begin{equation}
\begin{array}{l}
 \displaystyle
 \Ta = (-1)^j \frac{|1-\alpha r|^{\frac12}} 
 {(1+\alpha r)^{\frac{1-2\alpha m}{2(1+2\alpha m)}} 
 \left|1-\frac r{2m}\right|^{\frac{2\alpha m}{1+2\alpha m}}}
       \;\cosh\Bigl((1-2\alpha m)\,\alpha t\Bigr)\;, \\[15pt] 
 \displaystyle
 \Ra = (-1)^j \frac{|1-\alpha r|^{\frac12}}
 {(1+\alpha r)^{\frac{1-2\alpha m}{2(1+2\alpha m)}} 
 \left|1-\frac r{2m}\right|^{\frac{2\alpha m}{1+2\alpha m}}}
       \;\sinh\Bigl((1-2\alpha m)\,\alpha t\Bigr)\;.
\end{array} 
\end{equation}
 In the two domains of type~II, the corresponding expressions are identical except that the $\sinh$ and $\cosh$ terms are interchanged. After this transformation, the 2-space on which $\theta$ and $\varphi$ are constant has the metric 
 $$ \d s_2^2 ={2m\,(1+\alpha r)^{2\over1+2\alpha m} 
\left|1-{r\over2m}\right|^{1+6\alpha m\over1+2\alpha m}\over 
 \alpha^2(1-2\alpha m)^2\,r\,(1+\alpha r\cos\theta)^2}
\,\Big(-\d\Ta^2+\d\Ra^2\Big). $$
 Of course, $r$ must now be expressed in terms of $\Ta$ and $\Ra$ and this is smooth at \hbox{$r=1/\alpha$} for the above specific choice of the constant $k$. The metric is thus regular at the acceleration horizon, which now corresponds to the two lines $\Ta=\pm \Ra$.

We have thus extended the metric across the acceleration horizon and shown that the space-time contains not only one static domain of type~II (which corresponds to the exterior of a black hole), but {\em two} distinct static domains that are causally separated by the acceleration horizon $\Ha$, and also a second (prior) non-static domain of type~I.

Notice, however, that the range of the coordinates $\Ua$ and $\Va$ in the domains~I is restricted by the presence of conformal infinity. To cover this region fully, it is better to revert to the coordinates $x$ and $y$ using (\ref{xytothetar}) in which conformal infinity is given by $y=-x$, as this also permits $y$ to be negative. 
(Considering only positive values of $r$, $\scri$ is not reached for $\theta\in(0,{\pi\over2})$.) From the definitions (\ref{uvdef})--(\ref{UVuvtransf}) we see that this gives an upper limit (which depends on the value of ${\theta}$) on the product ${\Ua\Va}$.

We may similarly consider the space-time {\em near the black hole horizon}~$\Ho$ at $r=2m$, where we choose ${\k=\k_{\mathrm{o}}}$ so that ${\Uo=0}$ or ${\Vo=0}$ on the horizon. We now start with the static region of type~II with $(i,j)=(-1,0)$. By allowing ${\Uo,\,\Vo\in(-\infty,\infty)}$, we attach to it two other time-dependent domains of type~III (interior of the black and white holes) with $(-1,1),\,(-2,0)$, and an additional static domain of type~II with ${(-2,1)}$. The Kruskal--Szekeres-type coordinates (\ref{KruskCoor}) in the domains of type II are 
\begin{equation}
\begin{array}{l}
 \displaystyle
 \To = -(-1)^i\frac{\left|1-\frac r{2m}\right|^\frac12\, 
 (1+\alpha r)^{\frac{1-2\alpha m}{8\alpha m}}}
 {(1-\alpha r)^{\frac{1+2\alpha m}{8\alpha m}}}
       \;{\textstyle\sinh\Bigl(\frac{(1-4\alpha^2 m^2)}{4m}\, t\Bigr)}\;, \\[15pt] 
 \displaystyle
 \Ro = +(-1)^i\frac{\left|1-\frac r{2m}\right|^\frac12\,
 (1+\alpha r)^{\frac{1-2\alpha m}{8\alpha m}}}
 {(1-\alpha r)^{\frac{1+2\alpha m}{8\alpha m}}}
       \;{\textstyle\cosh\Bigl(\frac{(1-4\alpha^2 m^2)}{4m}\, t\Bigr)}\;,
\end{array} 
\end{equation}
and again the expressions for the domains of type III have identical forms with the $\sinh$ and $\cosh$ terms interchanged. In this case, the 2-space on which $\theta$ and $\varphi$ are constant has the metric 
 $$ \d s_2^2 ={32m^3\,(1-\alpha r)^{1+6\alpha m\over4\alpha m} \over
 (1-4\alpha^2m^2)\,r(1+\alpha r\cos\theta)^2 (1+\alpha r)^{1-6\alpha m\over4\alpha m} }
 \,\Big(-\d\To^2+\d\Ro^2\Big). $$ 
 This is clearly continuous at $r=2m$ and is the analogue of the Kruskal--Szekeres coordinates for the Schwarzschild black hole (to which it reduces in the limit $\alpha\to0$). In fact, the metric is smooth in these coordinates at ${\To=\pm\Ro}$, and we have thus obtained the extension of the metric across~$\Ho$. The domain covered by ${\Uo,\Vo}$ contains two static regions representing two black hole exteriors that are connected through the Einstein--Rosen bridge. The two domains of type~III correspond to the interiors of the black hole and the complementary white hole.
Notice in this case that the curvature singularity at ${r=0}$ in domains of type III is given by the condition ${\Uo\Vo=1}$.

\begin{figure}[h]
\begin{center}
\parbox[t]{300pt}{\centering\small
 \includegraphics{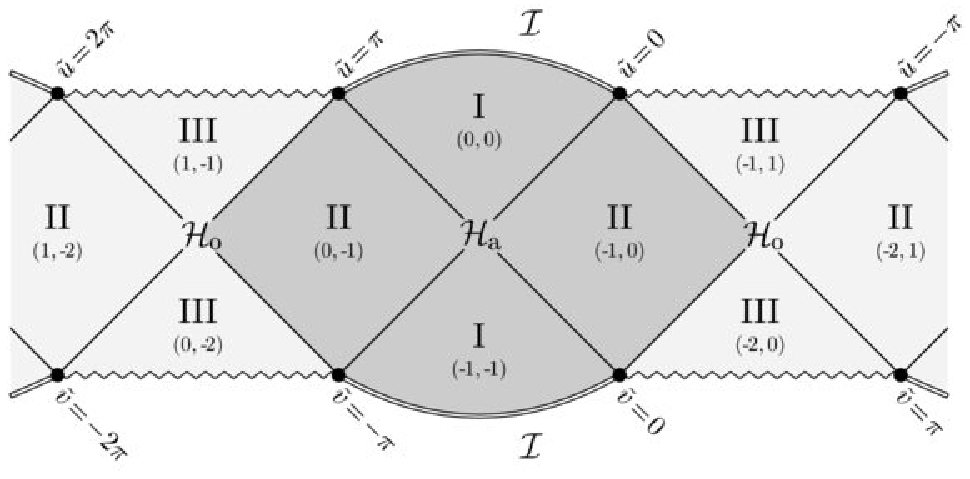}\\
 (a)}\\
\parbox[t]{150pt}{\centering\small
 \includegraphics[scale=0.8]{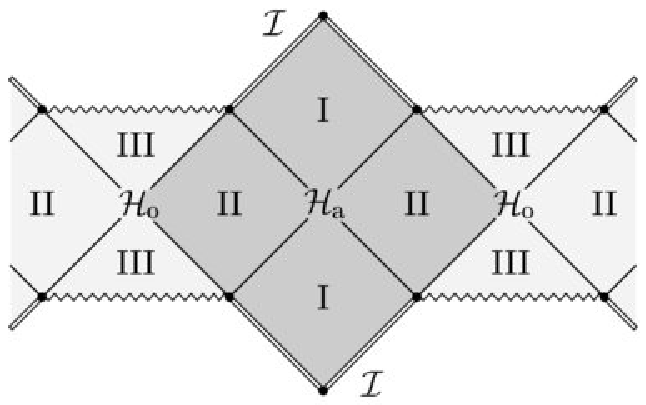}\\
 (b)}\quad
\parbox[t]{150pt}{\centering\small
 \includegraphics[scale=0.8]{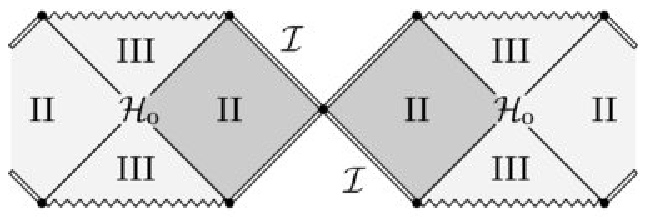}\\
 (c)}
\end{center}
\caption{\small 
Conformal diagrams for the complete space-time on sections on which ${\theta,\varphi=\hbox{const.}}$, for different values of~$\theta$. The entire space-time is composed of distinct domains of types I, II and III, labelled by the integers $(i,j)$. It is bounded by curvature singularities and by conformal infinity~${\cal I}$, as indicated in~(a). Diagrams (b) and (c) indicate the limiting cases on the axes $\theta=0$ and $\theta=\pi$ respectively. }
\label{fig:CmtrCD}
\end{figure}

In the same way, we can go across the horizon of the other black hole using another set of coordinates ${\Uo,\,\Vo}$ covering domains ${(i,j)=(0,-1),\,(0,-2),\,(1,-1)}$, and ${(1,-2)}$. This process can be repeated across any subsequent horizon and we can thus continue the space-time into a maximally extended manifold which contains an infinite sequence of domains. The sequence of coordinates ${\Ua,\,\Va}$ and ${\Uo,\,\Vo}$ each  cover smoothly one horizon and four adjacent blocks of the initial ${u,v}$-coordinates around it. The integers ${i,j}$ which label the different domains (see figure~\ref{fig:CmtrCD}) satisfy the constraint ${i+j=0,-1,-2}$.

Globally {\em compactified} coordinates ${\uE,\vE}$ which cover the whole space-time can be introduced by the definition
 \begin{equation}
 \label{utvtuvdef}
\begin{array}{l}
  \displaystyle\tan\frac\uE2
  =(-\sign\k)\;\, U^{-\sign\k}
  =(-1)^i\,\exp\Bigl(\frac{\alpha u}{2|\k|}\Bigr)\;,
  \qquad\uE\in\Big(i\pi,(i+1)\pi\Big)\;,\\[1.5ex]
  \displaystyle\tan\frac\vE2
  =(-\sign\k)\;\, V^{-\sign\k}
  =(-1)^j\,\exp\Bigl(\frac{\alpha v}{2|\k|}\Bigr)\;,
  \qquad\vE\in\Big(j\pi,(j+1)\pi\Big)\;.
\end{array}
 \end{equation}
Again, with the choices ${\k=\k_{\mathrm{a}}}$ or ${\k=\k_{\mathrm{o}}}$, the metric is smooth across the acceleration or black hole horizons respectively. On the horizons either of the coordinates ${\uE}$ or ${\vE}$ are integer multiples of ${\pi}$. Different domains of the coordinates ${u,v}$ thus corresponds to different blocks of the coordinates ${\uE,\,\vE}$ separated by horizons $\Ha$ and $\Ho$, which are naturally labeled by two integers ${(i,j)}$.
The whole space-time is covered by ${\uE,\vE\in(-\infty,\infty)}$, constrained by conditions at conformal infinity and the singularity. This is illustrated in the two-dimensional conformal diagrams given in figure~\ref{fig:CmtrCD}.

Finally, recall that the extension across the horizons and the compactification into two-dimensional conformal diagrams were obtained by transforming the coordinates ${t}$ and ${r}$ only: the angular coordinates ${\theta}$ and ${\ph}$ remained unchanged. However, the exact form of the conformal diagrams --- particularly, the location of~$\scri$ --- depends on the value of the coordinate~${\theta}$. For general ${\theta}$, the location of conformal infinity in the two-dimensional conformal diagram is given by a spacelike line as typically shown in figure~\ref{fig:CmtrCD}(a). It is not clear {\it a~priori} why conformal infinity is not manifestly null as one would expect for a vacuum solution of the Einstein equations without a cosmological constant. 
It may also be recalled that the coordinates $r,\theta$ do not reach as far as $\scri$ for $0<\theta<{\pi\over2}$. For this range, it is preferable to revert to the $x,y$-coordinates~(\ref{xytothetar}). With this, conformal infinity on the inner axes $x=1$ (${\theta=0}$) is seen to be null as shown in figure~\ref{fig:CmtrCD}(b). However, this axis is generally singular as will be discussed below. 
In addition, on the outer axis, ${\theta=\pi}$, conformal infinity ``coincides'' with the horizon at $r=1/\alpha$ as shown in figure~\ref{fig:CmtrCD}(c),
but this axis also generally corresponds to a topological singularity. These questions will be thoroughly discussed in the following sections.

\section{Some geometrical and physical properties}
\label{sc:PhysProp}
    The $C$-metric in either of the forms (\ref{HTCmetric}) or (\ref{BLCmetric}) is expressed in terms of {\em three} positive real parameters $m$ and $\alpha$ (satisfying ${2m\alpha<1}$) and ${C}$ (which is hidden in the range of the coordinate ${\ph\in(-\pi C,\pi C)}$). The basic interpretation of this solution is well know. The purpose of this section is to identify certain invariant geometrical and physical quantities and to express them in terms of these new parameters.

\begin{figure}[hpt]     
\centerline{
\includegraphics[scale=1.0]{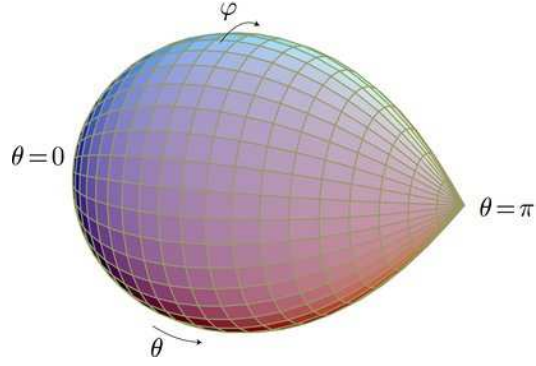} }
\caption{\small The surface of constant $t$ and $r$ illustrated as an embedding into E$^3$. This is regular at $\theta=0$, but there is a conical singularity at $\theta=\pi$ corresponding to the deficit angle $\delta_\pi={8\pi\alpha m\over1+2\alpha m}$. }
\label{tearfig}
\end{figure}

Let us start by examining the regularity of the axis of symmetry. For this, we consider a small circle around the half-axis $\theta=0$ (with $t,r$ const). For the above range of $\varphi$, we obtain
\begin{equation}\label{con0}
{\hbox{circumference}\over\hbox{radius}} 
 =\lim_{\theta\to0} {2\pi C P\sin\theta\over\theta} =2\pi C(1+2\alpha m)\;.
\end{equation}
In general, it is not exactly $2\pi$, which implies the existence of a conical singularity. Similarly, around the other half axis $\theta=\pi$, 
 \begin{equation}\label{conpi}
{\hbox{circumference}\over\hbox{radius}}
 =\lim_{\theta\to\pi} {2\pi C P\sin\theta\over\pi-\theta} 
 =2\pi C(1-2\alpha m)\;,
 \end{equation} 
 which implies the existence of a conical singularity with a \emph{different} conicity (unless ${\alpha m=0}$). The deficit or excess angles of either of these two conical singularities can be removed by an appropriate choice of the constant~${C}$, but not both simultaneously. In general, the constant~$C$ can thus be seen to determine the ballance between the deficit/excess angles on the two halves of the axis. In particular, one natural choice is to remove the conical singularity at $\theta=0$ by setting ${C=(1+2\alpha m)^{-1}}$. In this case,
the deficit angles at the poles $\theta=0,\pi$ are respectively 
 \begin{equation} 
 \delta_0=0 , \qquad \delta_\pi={8\pi\alpha m\over1+2\alpha m}\;. 
 \label{defangle} 
 \end{equation} 
 As first given by Kinnersley and Walker \cite{KinWal70}, a typical surface on which $t$ and $r$ are constant can be illustrated as in figure~\ref{tearfig} as a surface embedded in an artificial Euclidian space in such a way that it has the correct inner geometry induced by the metric. However, this is not the correct embedding in the real curved space-time as the coordinates $t$, $r$, $\theta$, $\ph$ of the $C$-metric (\ref{BLCmetric}) are orthogonal, so in the real space-time there is no sharp vertex.

The conical singularity with constant deficit angle along the half axis \hbox{$\theta=\pi$} can be interpreted as representing a {\em semi-infinite cosmic string} under tension. This extends from the source at $r=0$ right out to conformal infinity. This naturally leads to the interpretation that the metric (\ref{BLCmetric}) represents a Schwarzschild-like black hole that is being accelerated along the axis $\theta=\pi$ by the action of a force which corresponds to the tension in a cosmic string.

It is often assumed that the range of the rotational coordinate is $2\pi$. This can be achieved by the simple rescaling 
 $$ \varphi=C\,\phi, $$ 
 where $\phi\in(-\pi,\pi)$. For the above natural choice, where a cosmic string occurs only along the ${\theta=\pi}$ axis, the line element takes the form
 \begin{equation} 
 \d s^2={1\over(1+\alpha r\cos\theta)^2} 
 \left( -Q\,\d t^2 +{\d r^2\over Q} +{r^2\,\d\theta^2\over P}  
 +{ P\,r^2\sin^2\theta\over(1+2\alpha m)^2}\,\d\phi^2 \right) 
 \label{varBLCmetric} 
 \end{equation}  
where $P$ and $Q$ are still given by (\ref{BLPQ}).

The remaining metric parameters ${m}$ and ${\alpha}$ depend on the `gauge fixing' of the form of the ${C}$-metric. Rescaling coordinates can modify its form and lead to a different set of metric parameters. It is therefore useful to relate the parameters to invariant geometrical quantities. Here we list some of the physically most interesting ones.

The area of the black hole horizon ${r=2m}$ is given by 
\begin{equation}
 \label{BHarea}
  \mathcal{A} =\int_{-\pi C}^{\pi C} \int_0^\pi 
  \sqrt{g_{\theta\theta}\,g_{\ph\ph}} \>
\bigg|_{\!\!\begin{array}{l}\scriptscriptstyle r=2m\\[-1.2ex]\scriptscriptstyle  t=\hbox{\tiny const.}\end{array}}\!\!\d\theta\, \d\ph
  = \frac{16\pi C m^2}{1-4\alpha^2m^2}\;.
\end{equation}

The surface gravity of the horizons is defined by ${\kappa^2=-\frac12 \ell_{\mu;\nu}\ell^{\mu;\nu}}$, which is unique up to a normalization of the Killing vector ${\ell^\mu=\partial_t}$. An invariant quantity is thus given by the ratio of the surface gravity on the black hole and acceleration horizons. These are given by 
 \begin{equation}
 \label{surfgr}
 \kappa_{\mathrm{o}}={1-4\alpha^2m^2\over4m}, \qquad 
 \kappa_{\mathrm{a}}=\alpha(1-2\alpha m). 
 \end{equation} 
 which are respectively related to the Hawking and Unruh temperatures of the black hole and accelerating frame. It may also be observed that $\kappa_{\mathrm{o}}{\cal A}=4\pi Cm$ which, according to \cite{GerHar82}, implies that the mass of the black holes is given precisely by~$Cm$. Notice also that the ratio ${\kappa_{\mathrm{o}}/\kappa_{\mathrm{a}}}$ determines the quantity ${m\alpha}$, which can be understood as the force per unit angle ${\ph}$ acting on the holes. Indeed, it corresponds to a difference per unit angle in the tensions of the cosmic strings on the two parts of the axis, cf.\ eqs.~(\ref{con0}) and~(\ref{conpi}).

The distance between the points of bifurcation of the black hole and acceleration horizons along the axis ${\theta=0}$ at ${t=\hbox{constant}}$ is given by the complete elliptical integral
 \begin{equation}
 \label{hordist}
  \mathcal{L}=\int_{2m}^{1/\alpha}  \sqrt{g_{rr}} \>
  \bigg|_{\!\!\begin{array}{l}\scriptscriptstyle \theta=0\\[-1.2ex]\scriptscriptstyle  t=\hbox{\tiny const.}\end{array}}
  \!\!\d r
  =\frac1{\alpha\sqrt{1+2\alpha m}}\;\mathbf{E}\left(\frac{1-2\alpha m}{1+2\alpha m}\right).
 \end{equation}
 (Notice that this does not depend on the choice of $t$.)
 Similarly, the proper time from the point of bifurcation of the black hole horizon to the singularity along the axis ${\theta=0}$ and ${t=\hbox{constant}}$ is given by 
 \begin{equation}
 \label{horsingtime}
  \mathcal{T}=\int_0^{2m}\!\! \sqrt{-g_{rr}} \> 
  \bigg|_{\!\!\begin{array}{l}\scriptscriptstyle \theta=0\\[-1.2ex]\scriptscriptstyle  t=\hbox{\tiny const.}\end{array}}
  \!\!\d r
  =\frac1{\alpha\sqrt{1+2\alpha m}}
  \left[\mathbf{K}\left(\frac{4\alpha m}{1+2\alpha m}\right)
  -\mathbf{E}\left(\frac{4\alpha m}{1+2\alpha m}\right) \right],
 \end{equation} 
 which is finite and reduces to $\pi m$ when $\alpha=0$.

\section{The weak field limit}
\label{sc:WeakField}
     In order to understand the meaning of the parameter $\alpha$, it is natural to consider the weak field limit $m\to0$. In this case, the metric (\ref{BLCmetric}) becomes 
 \begin{equation} 
 \begin{array}{l}
 \d s^2
 \displaystyle={1\over(1+\alpha r\cos\theta)^2} \left[ -(1-\alpha^2r^2)\,\d t^2 
 +{\d r^2\over(1-\alpha^2r^2)} +r^2(\d\theta^2 +\sin^2\theta\,\d\ph^2) \right]. 
 \label{MinkClimit} 
 \end{array}
 \end{equation}  
Since the curvature tensor (\ref{CWeyl}) vanishes in this limit, this metric must represent (at least a part of) Minkowski space. In fact, we can apply the transformation  
\begin{equation}  
  \zBP=\frac{\sqrt{|1-\alpha^2r^2}|}{\alpha(1+\alpha r\cos\theta)}
 = \frac{\sqrt{|y^2-1|}}{\alpha(x+y)}\;, \qquad
  \rBP={r\sin\theta\over1+\alpha r\cos\theta}
  =\frac{\sqrt{1-x^2}}{\alpha(x+y)} \;, 
 \label{BPcoor}
\end{equation}
 with ${\tau=\alpha t}$, leaving ${\ph}$ unchanged. With this, the metric (\ref{MinkClimit}), in the region~II in which ${r<1/\alpha}$, takes the (cylindrical) uniformly accelerating Rindler form 
 \begin{equation} 
 \d s^2= -\zBP^2\,\d \tau^2 +\d \zBP^2 +\d\rBP^2 +\rBP^2\d\ph^2\,. 
 \label{RindlMink} 
 \end{equation}
This can be put into the standard cylindrical form \ ${-\d s^2=\d \TBP^2-\d \ZBP^2-\d\rBP^2-\rBP^2\d\ph^2}$ \ of the Minkowski metric by the transformation
 ${\TBP=\pm\zBP\sinh\tau}$, ${\ZBP=\pm\zBP\cosh\tau}$,
 where $\zBP\in(0,\infty)$ and $\tau\in(-\infty,\infty)$ and the choice of signs (giving two copies of the region~II) in both expressions is the same.  
In the region~I in which ${r>1/\alpha}$, the metric (\ref{MinkClimit}) can be similarly transformed using (\ref{BPcoor}) followed by
${\TBP=\pm\zBP\cosh\tau}$, ${\ZBP=\pm\zBP\sinh\tau}$, 
where again the choice of signs is the same for the two copies of this region.

\begin{figure}[hpt]     
\begin{center}
\parbox[b]{200pt}{\centering\small
 \includegraphics[scale=0.4, trim=5 5 5 15]{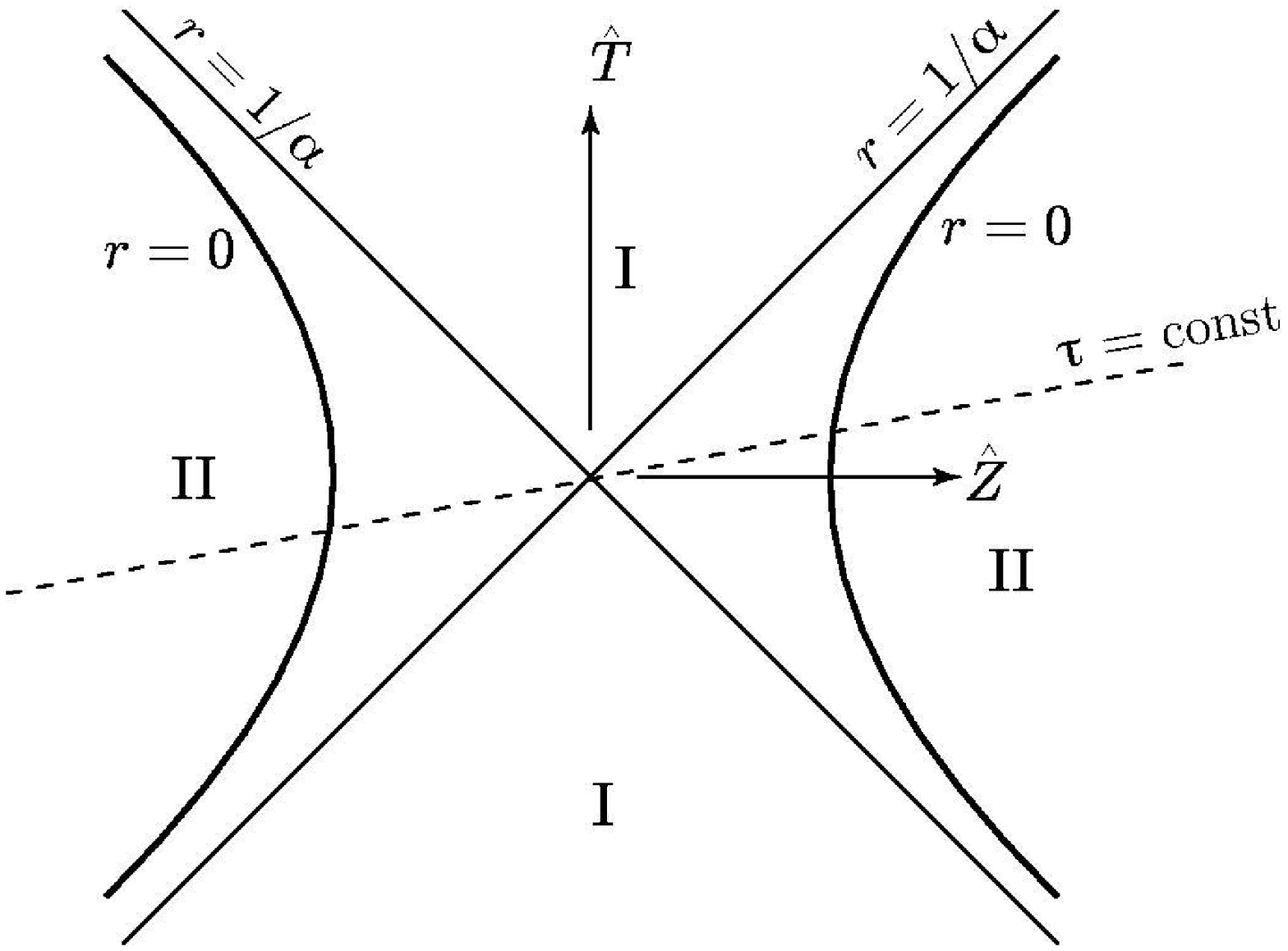}\\[2ex]
 (a)}\qquad\quad
\parbox[b]{200pt}{\centering\small
 \includegraphics{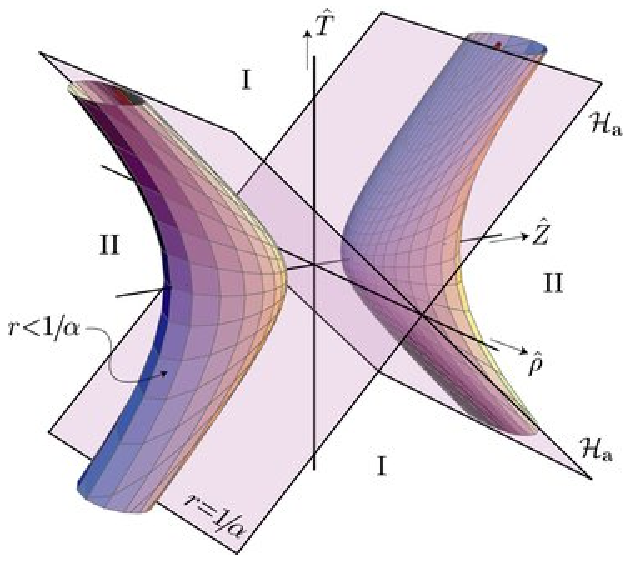}\\
 (b)}
\end{center}
\caption{\small In the weak field limit of the $C$-metric, a causally separated pair of test particles at $r=0$ accelerate away from each other in opposite spatial directions as illustrated in (a). The acceleration horizon at $r=1/\alpha$ is given by $\ZBP=\pm \TBP$. For ${r<1/\alpha}$, surfaces ${r=\hbox{const.}}$  enwrap  the accelerated particles at ${r=0}$ as depicted in (b).}
\label{fig:Mink}
\end{figure}

The complete Minkowski space is thus covered by four regions, of either type I or II, that are separated by the null hypersurfaces $\ZBP=\pm\TBP$ on which $r=1/\alpha$. These are acceleration horizons which are Killing horizons forming the boundaries of the regions described by the uniformly accelerating (Rindler) metric (\ref{RindlMink}).  
In either of the regions~II, points with constant values of $r$, $\theta$ and $\ph$ have uniform acceleration: i.e., they follow the worldline 
$\ZBP^2-\TBP^2={1-\alpha^2r^2\over \alpha^2(1+\alpha r\cos\theta)^4}$. 
In particular, {\em the acceleration of test particles located at the origins $r=0$ is exactly $\alpha$} in the positive or negative $\ZBP$ direction, according to the sign of $\ZBP$ as illustrated in figure~\ref{fig:Mink}(a). This provides a clear physical meaning for the parameter $\alpha$ in the weak field regime.

Let us now consider a smooth spatial section $\tau=$~const. through the regions~II of the above Minkowski space. This passes through $\TBP=0=\ZBP$, which is where the acceleration horizon bifurcates. Each half section, $\ZBP>0$ or $\ZBP<0$, is covered by the coordinate set $r,\theta,\ph$ with $r\in(0,1/\alpha)$. To cover the complete spatial section, it could be convenient to introduce a new coordinate $\eta$ defined such that $\cosh\eta=1/\alpha r$, where $\eta\in(-\infty,0)$ for ${\ZBP<0}$ and $\eta\in(0,\infty)$ for ${\ZBP>0}$.

\begin{figure}[hpt]     
\centerline{
\includegraphics{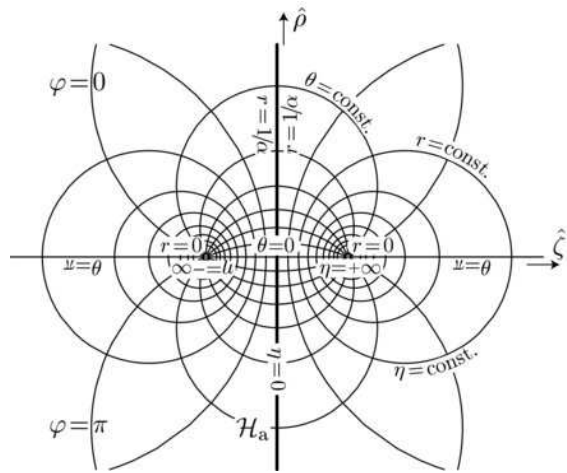} }
\caption{\small
The coordinate lines of ${\{r,\theta\}}$ (which are equivalent to the lines of ${\{\eta,\theta\}}$ or ${\{y,x\}}$) drawn with respect to flat background space ${\{\zBP,\rBP\}}$ on a spatial section of Minkowski space with constant $\tau$ and which passes through $\TBP=\ZBP=0$ which is the line of bifurcation of the acceleration horizons. This clearly demonstrates the bi-spherical structure of these coordinates. }
\label{fig:BPMink}
\end{figure}

Obviously, the coordinate pairs ${\{r,\,\theta\}}$, ${\{\eta,\,\theta\}}$ and ${\{y,x\}}$ all have the same coordinate lines. In the plane ${\tau=\hbox{const.}}$, these are drawn (with ${\ph}$ ignored) in figure~\ref{fig:BPMink} using axes which correspond to Rindler coordinates ${\zBP}$ and ${\rBP}$ of (\ref{BPcoor}). We see that these lines have \emph{bi-polar structure} (or bi-spherical structure, if we include the angular coordinate ${\ph}$ around the axis $\rBP=0$) with origins at ${r=0}$. The coordinates ${r,\,\theta,\,\ph}$ thus play the role of spherical coordinates near the origins, i.e., around the accelerated test particles, cf.\ figure~\ref{fig:Mink}(b). However, these coordinates deform as $r\to1/\alpha$.

\section{Global structure in the weak field limit} 
\label{sc:glstrWFL}
      When $m=0$, the $C$-metric reduces to flat Minkowski space whose global structure is well understood and can be visualised in many ways. However, as will now be shown, the conformal diagrams with ${\theta,\ph=\hbox{const.}}$ constructed in the same way as in section~\ref{sc:horizons} starting with the metric (\ref{MinkClimit}) are not the familiar two-dimensional conformal diagrams. 
In this case, the expression for the coordinate ${r_\star}$ (\ref{tortr}) can be explicitly inverted to give\footnote{\label{ftn:negr}
     To cover the complete region~I up to $\scri$ for all $\theta$, we may revert to the $y$ coordinate which can be negative. } 
\begin{equation}  
  \alpha r = \bigg\{
   \begin{array}{ll}
   \coth\alpha r_\star\qquad &\hbox{in the region I}\;, \\[0.5ex]
   \tanh\alpha r_\star\qquad &\hbox{in the region II}\;.
   \end{array}
\end{equation}
Double null coordinates ${u,v}$ are defined again by (\ref{uvdef}) and their compactified versions ${\uE,\,\vE}$ are given by (\ref{utvtuvdef}) with $2|k|=1$. The whole space-time is covered by two domains of type I and two domains of type II labeled by ${(i,j)=(0,0),(-1,-1)}$ and ${(i,j)=(-1,0),(0,-1)}$, respectively. These are separated by the horizon ${\Ha}$ at ${\uE=0}$ or ${\vE=0}$ corresponding to $r=1/\alpha$. In the domains II, the coordinate range is restricted by the condition that ${r>0}$. The value ${r=0}$ represents the trajectory of the test particles corresponding to the limit of the black holes, rather than a physical singularity.

\begin{figure}[h]
\begin{center}
\parbox[t]{60pt}{\centering\small
 \includegraphics{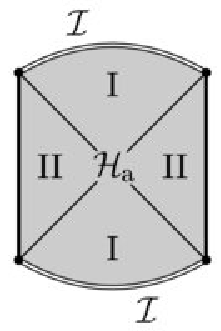}\\
 (a)}\qquad\quad
\parbox[t]{60pt}{\centering\small
 \includegraphics{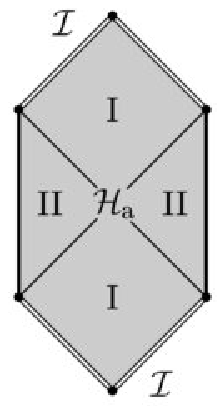}\\
 (b)}\qquad\quad
\parbox[t]{60pt}{\centering\small
 \includegraphics{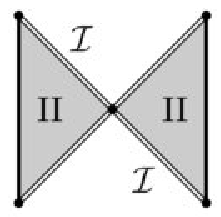}\\
 (c)}
\end{center}
\vskip-15pt
\caption{\small Conformal diagrams of typical sections ${\theta,\ph=\hbox{const.}}$ in the weak field limit,with axes given by coordinates ${\tE}$ and ${\tilde\chi}$ 
(and with null coordinates ${\tilde u,\,\tilde v}$ in diagonal directions) for (a) general $\theta$, (b) $\theta=0$, and (c) $\theta=\pi$. The accelerating test particles at $r=0$ are located on the vertical boundaries of region~II, and conformal infinity is illustrated by the double line boundaries of regions~I. } 
\label{fig:MinkCmtrCD}
\end{figure}

Introducing the coordinates ${\uE=\tE+\tilde\chi}$, ${\vE=\tE-\tilde\chi}$, the above transformations combine in both regions~I and II to give 
 \begin{equation}
 \label{xy-rotEinsTr}
  \alpha t=\frac12\log\left|\frac{\sin\tE-\sin\tilde\chi}{\sin\tE+\sin\tilde\chi}\right|\;,\qquad
  \alpha r = \frac{\cos\tilde\chi}{\cos\tE}\;, 
\end{equation}
 which convert the metric (\ref{MinkClimit}) to the form
 \begin{equation} 
 \label{EinstRotMtrc}
  \d s^2 = \frac{1}{\alpha^2(\cos\tE+\cos\tilde\chi\cos\theta)^2}
  \Bigl[ -\d\tE^2 +\d\tilde\chi^2
  +\cos^2\tilde\chi\,\bigl(\d\theta^2+\sin^2\theta\,\d\ph^2\bigr)\Bigr]\;.
 \end{equation}
 This is explicitly conformally related to the metric of the Einstein static universe with conformal factor ${\Omega=\cos\tE+\cos\tilde\chi\cos\theta}$. Conformal infinity occurs where $\Omega=0$, and this obviously depends on the choice of $\theta$. Two-dimensional conformal diagrams on which $\theta$ and $\varphi$ are constant are illustrated in figure~\ref{fig:MinkCmtrCD}. These are unfamiliar figures which will be shown to correspond to particular sections of the familiar three-dimensional (double cone) representation of the conformal Minkowski space.

The non-standard form of the compactified Minkowski space (\ref{EinstRotMtrc}) is related to the familiar one: 
 \begin{equation} 
 \label{EinstMtrc}
  \d s^2 = \frac{1}{\alpha^2(\cos\tE+\cos\chi)^2}\Bigl[
  -\d\tE^2 +\d\chi^2 +\sin^2\chi\,\bigl(\d\tht^2+\sin^2\tht\,\d\ph^2\bigr)\Bigr]\;,
 \end{equation}
 in which the conformal factor ${\cos\tE+\cos\chi}$ is independent of $\tht$, via the transformation 
 \begin{equation}
\label{rotEints-EinstTr}
 \begin{array}{l}
 \sin\tilde\chi =\sin\chi\cos\vartheta, \\[5pt]
 \tan\theta=-\tan\chi\sin\vartheta.
 \end{array} \hskip5pc
 \begin{array}{l}
 \cos\chi =\cos\tilde\chi\cos\theta, \\[5pt]
 \tan\vartheta=-\cot\tilde\chi\sin\theta.
 \end{array} 
 \end{equation} 
 Let us note that the metric (\ref{EinstMtrc}) can be directly obtained from the Minkowski metric in spherical coordinates by applying the transformation 
 \begin{equation}
 \label{SphMink-EinstTr}
 \alpha\TBP=\frac{\sin\tE}{\cos\tE+\cos\chi}\;, \qquad 
 \alpha\RBP=\frac{\sin\chi}{\cos\tE+\cos\chi}\;.
 \end{equation} 
 The coordinates ${\tE,\,\chi}$ are those used for the construction of the usual Minkowski conformal diagrams in which conformal boundary~$\scri$ is manifestly null.

\begin{figure}[h,t]
\begin{center}
\parbox[t]{200pt}{\centering\small
 \includegraphics{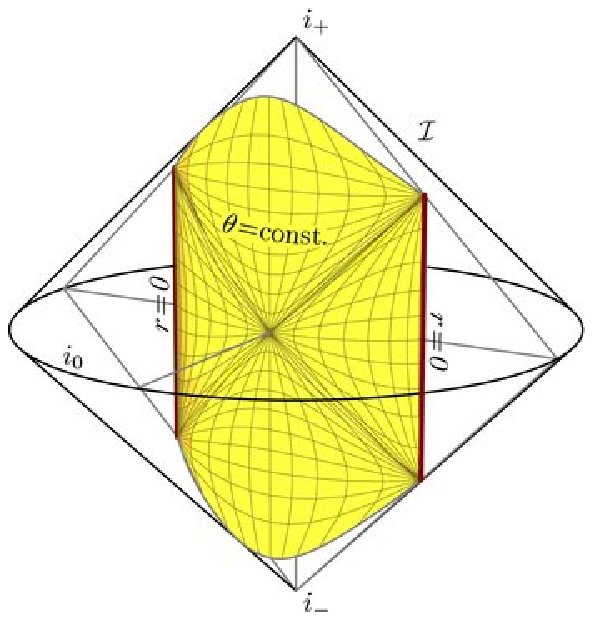}\\
 (a)}\qquad\quad
\parbox[t]{200pt}{\centering\small
 \includegraphics{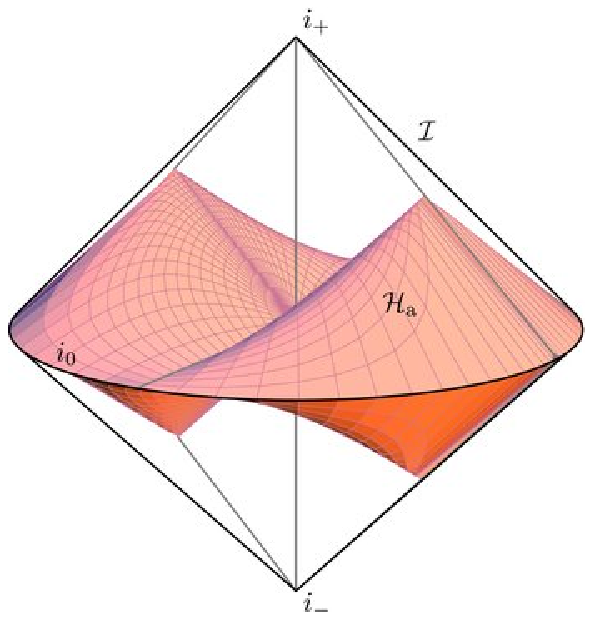}\\
 (b)}\end{center}
\vskip-12pt
\caption{\small (a) The conformal diagram of figure~\ref{fig:MinkCmtrCD}(a), and (b) the acceleration horizon, are illustrated as specific sections of the familiar double cone representation of conformally compactified Minkowski space. The points $i_-$ and $i_+$ denote past and future timelike infinity, while $i_0$ denotes spacelike infinity. }
\label{MinkSections}
\end{figure}

The non-standard diagrams in figure~\ref{fig:MinkCmtrCD} arise because we started with the metric~(\ref{MinkClimit}) whose bipolar coordinates ${r,\theta}$ are adapted to the pair of uniformly accelerating test particles. However, we can use the standard compactified coordinates (\ref{EinstMtrc}) and the relation (\ref{rotEints-EinstTr}) as a tool to visualize the two-dimensional surfaces $\theta,\ph=$~const. of figure~\ref{fig:MinkCmtrCD} in the familiar three-dimensional ``double cone'' diagram of the Minkowski space. Such a global picture is illustrated in figure~\ref{MinkSections}.

Suppressing $\ph$, the spatial section $\tE=0$ ($\tau=0$) through the two regions of type~II is now a `compactified' version of the surface with bipolar coordinates illustrated in figure~\ref{fig:BPMink}. As argued above, the accelerating test particles at $r=0$ are represented by two vertical lines in the conformal diagrams and thus they are also vertical in figure~\ref{MinkSections}. Similarly, the surfaces ${\theta,\ph=\hbox{const.}}$, corresponding to the family of conformal diagrams in figure~\ref{fig:MinkCmtrCD}, can be represented as vertical cylindrical surfaces (since the transformations (\ref{rotEints-EinstTr}) are time independent) spanned between the worldlines of the origins ${r=0}$. This clarifies the apparent paradox that conformal infinity~$\scri$ in the two-dimensional conformal diagrams in figure~\ref{fig:MinkCmtrCD} generally have a ``spatial'' character: they occur as \emph{intersections} of the time-like cylindrical surface ${\theta,\ph=\hbox{const.}}$ with the conical null conformal infinity of Minkowski space.

\section{Global structure of the $C$-metric} 
\label{sc:glstr}
     Let us now return to the full ${C}$-metric with ${m\neq0}$. The original metric (\ref{BLCmetric}) has already been extended across the horizons in section~\ref{sc:horizons}, and conformal diagrams were constructed. Our purpose here is to combine the two-dimensional sections in figure~\ref{fig:CmtrCD} to construct a helpful three-dimensional representation of the space-time (with the trivial ${\ph}$ direction suppressed) ouside the black hole horizons. In order to smoothly join these conformal diagrams for different ${\theta}$, we use the insights gained in the previous section.

By comparing the conformal diagrams in figure~\ref{fig:CmtrCD} with those of the corresponding weak field limit in figure~\ref{fig:MinkCmtrCD}, it may immediately be observed that the asymptotic structure of region~I is similar in the two cases. Thus, the asymptotic structure of the $C$-metric space-time close to $\scri$ must be basically the same as that illustrated in figure~\ref{MinkSections}. Differences will occur, however, with the introduction of the sources.

From the diagrams in figure~\ref{fig:CmtrCD}, it can be seen that each asymptotic domain of type~I extends to two black holes. Each of these has an internal structure which is 
qualitatively similar to that of the familiar Schwarzschild black hole. These black holes are causally separated by the acceleration horizon~$\Ha$. The inner structure of a single black hole is well understood, and we will therefore consider only the exterior of the black holes outside their horizons~$\Ho$. Accordingly, we will concentrate here on a combination of the domains of type~I, which contain conformal infinity, and the static domains of type~II. One such external region is illustrated by the darker shaded regions in the conformal diagrams of figure~\ref{fig:CmtrCD}.

\begin{figure}[h]
\begin{center}
\parbox[b]{200pt}{\centering\small
 \includegraphics[scale=1.25]{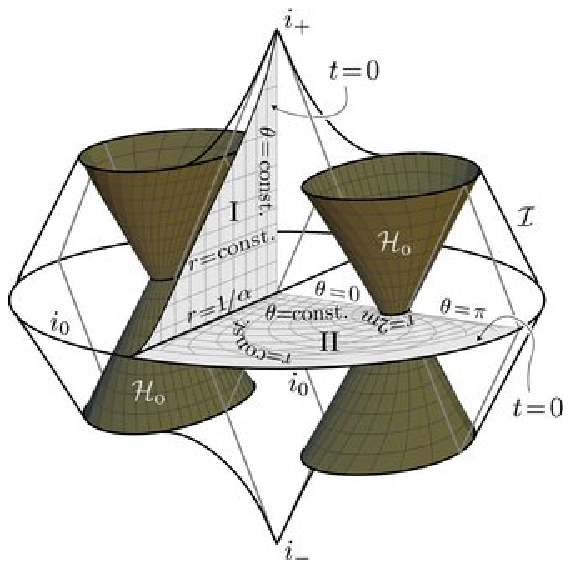}\\[1ex]
 (a)}
\parbox[b]{200pt}{\centering\small
 \includegraphics{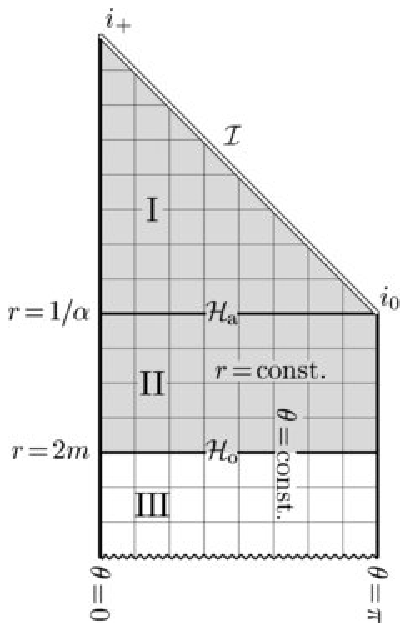}\\[1ex]
 (b)}
\end{center}
\vskip-1.3pc
\caption{\small The framework for constructing a representation of the global structure of the $C$-metric space-time outside the black hole horizons (i.e. regions of types I and II) is given in (a) with the ${\ph}$ direction suppressed. The corresponding part of figure~\ref{PDcoords} is shown as the shaded region in (b). The coordinate lines on the surface $t=0$ are shown in both (a) and~(b).   }
\label{fig:3Dxyembed}
\end{figure}

To construct a representation of the causal structure of the space-time, we start -- as shown in figure~\ref{fig:3Dxyembed}(a) -- by considering the spacelike surface ${t=0}$ in regions~II, on which the $r,\theta$ coordinates have bipolar structure as in figures~\ref{fig:BPMink}. However, we now have to replace the two point-like origins ${r=0}$ by two circles ${r=2m}$, which each correspond to bifurcations of the black hole horizons~$\Ho$. Next, we need to include a timelike dimension in the vertical direction.

\begin{figure}[h]
\begin{center}
\parbox[b]{200pt}{\centering\small
 \includegraphics{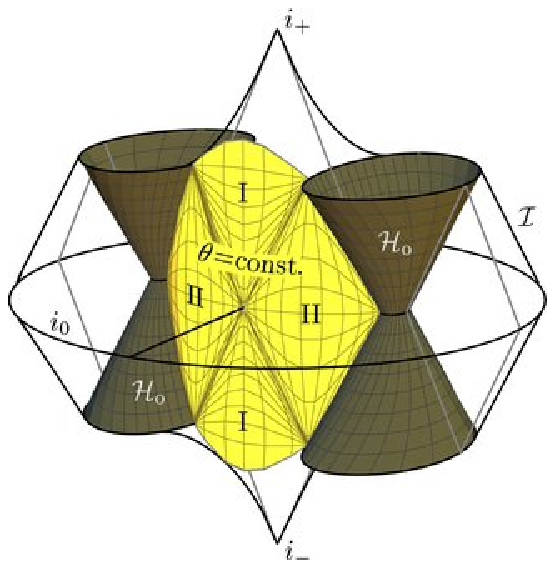}\\[1ex]
 (a)}\\[-8ex]
\parbox[b]{200pt}{\centering\small
 \includegraphics{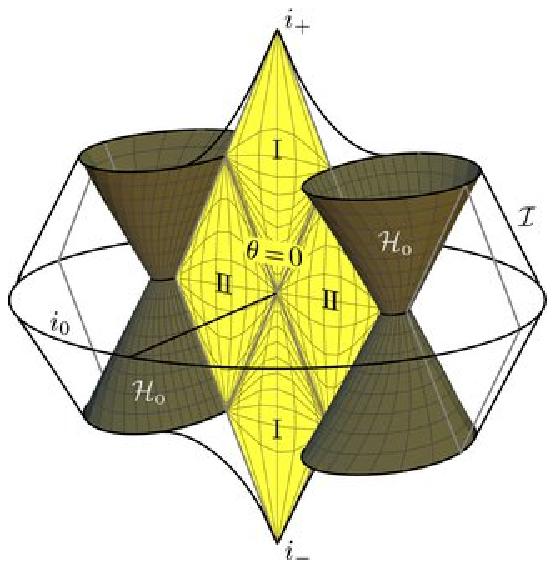}\\[1ex]
 (b)}
\parbox[b]{200pt}{\centering\small
 \includegraphics{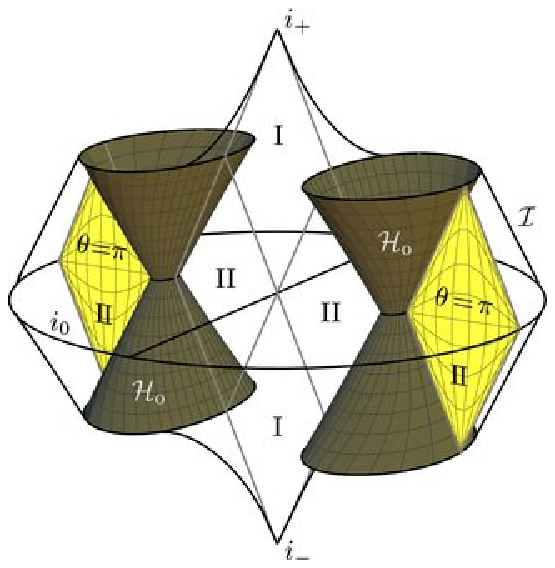}\\[1ex]
 (c)}
\end{center}
\vskip-1.3pc
\caption{\small Vertical cylindrical surfaces on which $\theta=$~const. are the same sections as the conformal diagrams in figure~\ref{fig:CmtrCD}. These surfaces can be drawn for ${\theta\in(0,\pi)}$ and two antipodal values of ${\ph}$ (e.g., ${\ph=0,\pi}$). Together, they lead to a three-dimensional representation of the conformally compactified $C$-metric space-time, which shows its global structure outside the black hole horizons. The special cases with $\theta=0,\pi$, which generally correspond to conical singularities, are included in (b) and (c). }
\label{fig:3DCD}
\end{figure}

The global structure of the weak field limit of the $C$-metric was illustrated in figure~\ref{MinkSections}. In that figure, each vertical cylindrical surface $\theta=$~const. corresponds to a two-dimensional conformal diagram of the family pictured in~figure~\ref{fig:MinkCmtrCD}. In a similar way, we now construct an analogous figure describing the global structure of the full $C$-metric outside the black hole horizons.  This is achieved by raising cylindrical surfaces vertically above the lines ${\theta=\hbox{const.}}$ in the plane $t=0$. On these surfaces, we draw the corresponding conformal diagrams given in figure~\ref{fig:CmtrCD} for the same constant~$\theta$. Typical examples are shown\footnote{
    For further pictures and animations see \cite{webpage} } 
    in figure~\ref{fig:3DCD}.

\begin{figure}[ht]
\begin{center}
\parbox[b]{200pt}{\centering\small
 \includegraphics{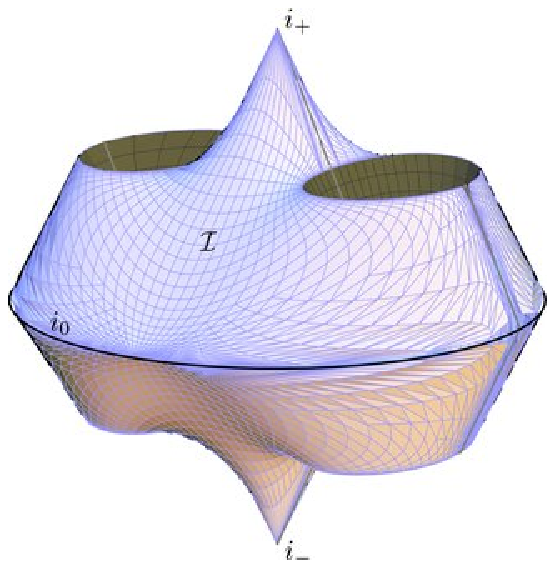}\\[1ex]
 (a)}\\[-8ex]
\parbox[b]{200pt}{\centering\small
 \includegraphics{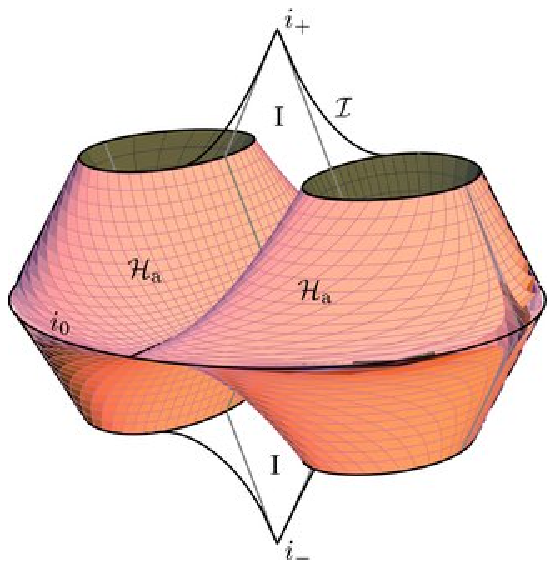}\\[1ex]
 (b)}
\parbox[b]{200pt}{\centering\small
 \includegraphics{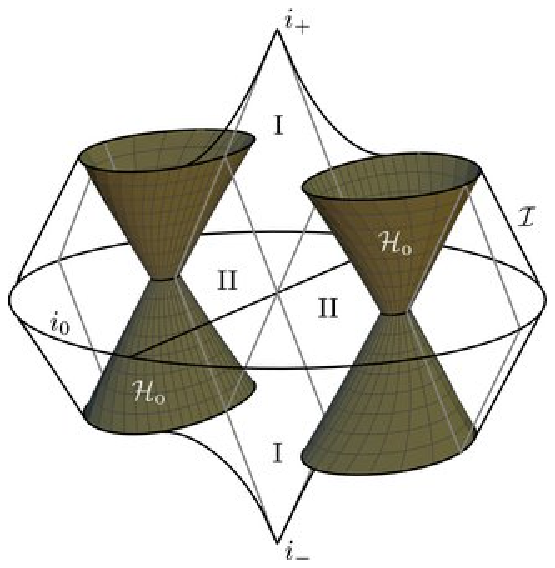}\\[1ex]
 (c)}
\end{center}
\vskip-1.3pc
\caption{\small Representaions of (a) conformal infinity~${\scri}$, (b) the acceleration horizon~$\Ha$, and (c)~the two black hole horizons~$\Ho$. }
\label{fig:3Dcaus}
\end{figure}

By taking the complete family of such surfaces for all $\theta\in(0,\pi)$, we construct an illustrative three-dimensional representation of the conformally compactified $C$-metric space-time which captures its global properties and causal structure. Moreover, this construction explicitly identifies coordinate labels at all points. This not only allows various coordinate surfaces to be drawn, but also identifies the physically significant ones. Specifically, surfaces representing conformal infinity~$\scri$, the acceleration horizon~$\Ha$ and the black hole horizons~$\Ho$ are pictured in figure~\ref{fig:3Dcaus}.

The figure~\ref{fig:3Dcaus}(a) shows the conformal infinity of the space-time. It is a null surface which resembles Minkowski-like $\scri$, but with two pairs of ``holes'' which represent the locations where the black hole horizons~$\Ho$ intersect conformal infinity. 
The figure~\ref{fig:3Dcaus}(b) illustrates the structure of the acceleration horizon~$\Ha$. We see that this horizon ``touches'' infinity only in two pairs of null generators on the axis $\theta=\pi$ as in figure~\ref{fig:CmtrCD}(c). It also completely encloses the static regions of type~II around the two black holes. The black hole horizons~$\Ho$, shown in figure~\ref{fig:3Dcaus}(c) are two null cones. They are generalizations of the vertical origins ${r=0}$ in figure~\ref{MinkSections} which describes the weak field limit.

\begin{figure}[h]
\begin{center}
\parbox[b]{200pt}{\centering\small
 \includegraphics{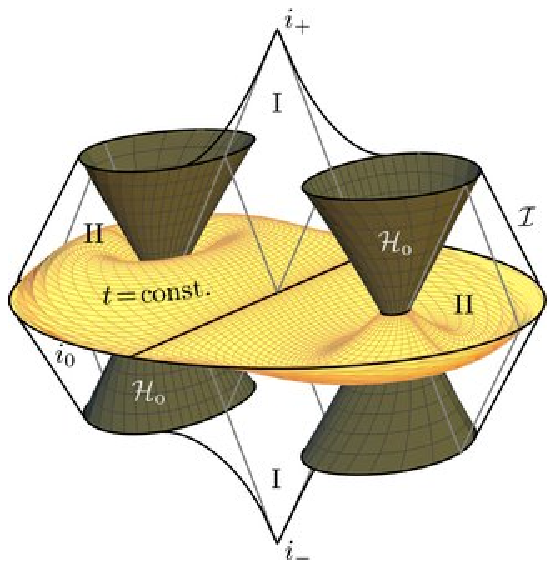}}
\parbox[b]{200pt}{\centering\small
 \includegraphics{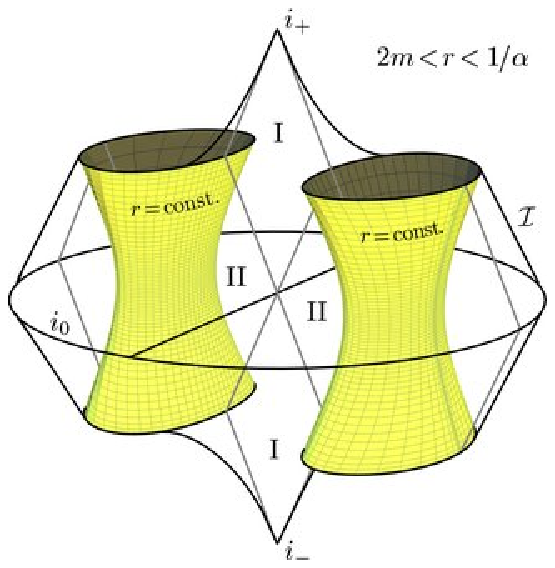}}\\[12pt]
\parbox[b]{200pt}{\centering\small
 \includegraphics{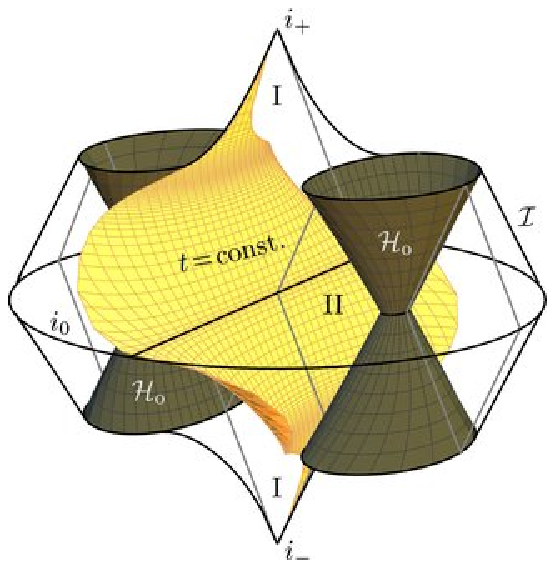}}
\parbox[b]{200pt}{\centering\small
 \includegraphics{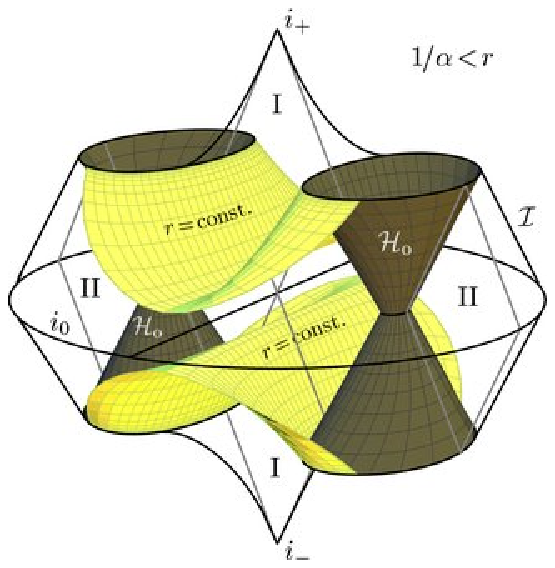}}
\end{center}
\vskip-1.3pc
\caption{\small The coordinate surfaces ${t,r=\hbox{const.}}$ in regions~II (above) and regions~I (below). }
\label{fig:3Drcoor}
\end{figure}

Finally, the embedding of the typical coordinate surfaces ${t=\hbox{const.}}$ and ${r=\hbox{const.}}$ into all depicted domains I and II is shown in figure~\ref{fig:3Drcoor}. Note that the initial particular surface ${t=0}$ was illustrated in figure~\ref{fig:3Dxyembed}(a).

\newpage
\section{Boost-rotation symmetric form} 
\label{sc:BoostRot}
\begin{figure}[ht] 
\centerline{
\parbox[t]{200pt}{\centering\small\includegraphics{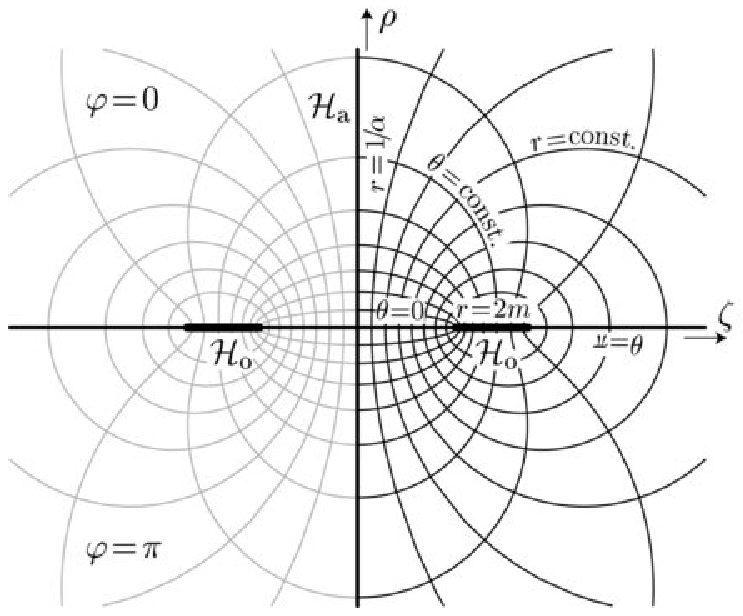} \\(a)}\qquad
\parbox[t]{200pt}{\centering\small\includegraphics{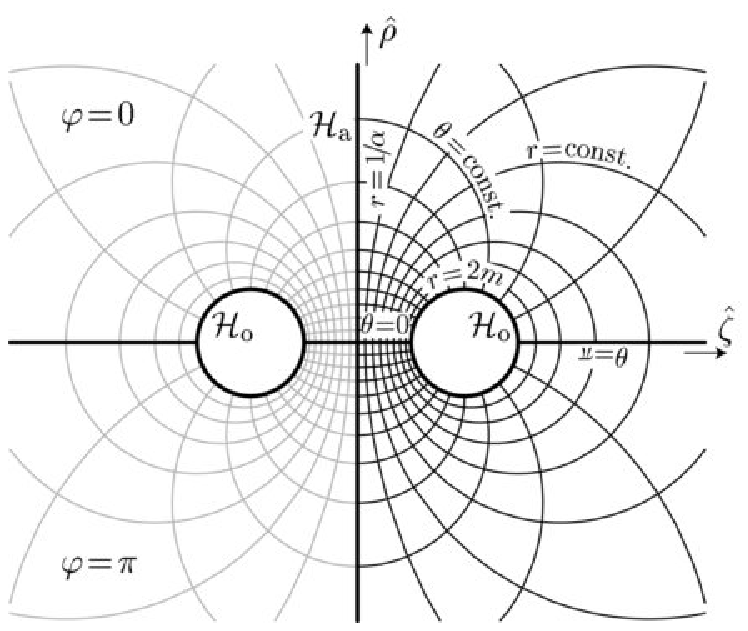}\\(b)}}
\caption{
\small The $r,\theta$-coordinate lines are plotted on the section $\tau=$~const. through two static regions~II with respect to (a) a boost-rotation symmetric grid ${\{\zBR,\rho\}}$ given by~(\ref{brtrans}), and (b) a space with grid ${\{\zBP,\rBP\}}$ given by (\ref{BPcoor}), which is part of figure~\ref{fig:BPMink} with the coordinate~$r$ bounded by two circles representing the black hole horizons~$\Ho$. In (a), these horizons are contracted (due to the factor $\sqrt{1-\frac{2m}{r}}$ in (\ref{brtrans})) to two abscissae located symmetrically on the axis ${\rho=0}$. The lines ${\zBR=0}$ and ${\zBP=0}$ represent the bifurcation of the acceleration horizon~$\Ha$. }
\label{fig:CM}
\end{figure}

     As another argument indicating that the space-time represents a pair of accelerated sources, we will now put the $C$-metric in a form in which the boost and rotation symmetries are explicitly manifested. To achieve this, we apply the transformation (which generalises (\ref{BPcoor})): 
 \begin{equation} 
 \zBR= \frac{\sqrt{|1-\alpha^2r^2|}}{\alpha(1+\alpha r\cos\theta)}\,
 \sqrt{1+2\alpha m\cos\theta}, \qquad
 \rho= \frac{r\,\sin\theta}{1+\alpha r\cos\theta}\,\sqrt{1-\frac{2m}{r}},
 \label{brtrans} 
 \end{equation}  
 with ${\tau=\alpha t}$, and ${\varphi}$ unchanged. A visualization of the relation between these coordinates is shown in figure~\ref{fig:CM}(a). With respect to the ${\{\zBR,\rho\}}$ grid, the ${r,\theta}$-coordinates have a ``bi-elliptical'' character, with the black hole horizons degenerated to two intervals on the axis ${\rho=0}$. Indeed, by inspecting (\ref{brtrans}) we immediately observe the correspondence: 
$$\begin{array}{llll} 
 \hbox{Acceleration horizon}\ \Ha:\ 
 &{r=\frac1\alpha}, 
 & \leftrightarrow\ 
 & \zBR=0,\ 0<\rho<\infty,  \\[1ex]
 \hbox{Inner axis:} 
 & \theta=0, 
 &  \leftrightarrow\ 
 &  \rho=0,\ 0< \zBR \le \frac1\alpha \sqrt{1-2\alpha m}, \\[1ex] 
 \hbox{Black hole horizon}\ \Ho:
 & r=2m, 
 & \leftrightarrow\ 
 & \rho=0,\ \frac1\alpha\sqrt{1-2\alpha m} \le \zBR\le \frac1\alpha\sqrt{1+2\alpha m}, \\[1ex] 
 \hbox{Outer axis:} 
  & \theta=\pi, 
  & \leftrightarrow\  
  & \rho=0,\ \frac1\alpha\sqrt{1+2\alpha m}\le \zBR\, .
 \end{array} $$ 
 For comparison, the corresponding picture for the similar grid given by~(\ref{BPcoor}) is given in figure~\ref{fig:CM}(b). 
Note that the coordinates $\{r,\theta\}$, $\{\zBR,\rho\}$, and $\{\zBP,\rBP\}$ cover only (the right) half of the section $\tau=\hbox{const.}$ To vizualize these coordinates around both holes, another copy of these coordinates is drawn in grey in the other (left) half of the figures. Two antipodal values of the coordinate $\ph$ have also been included.

Using (\ref{brtrans}) in region~II, the metric (\ref{BLCmetric}) takes the form 
 \begin{equation}
 \label{brdiag-metric}
  \d s^2 = -e^\mu\zBR^2\d\tau^2 +e^\lambda\bigl(\d\zBR^2+\d\rho^2\bigr) 
  +e^{-\mu}\rho^2\d\varphi^2\;,
 \end{equation}
 where the functions ${\mu}$ and ${\lambda}$ are given by
  \begin{equation} 
 \begin{array}{l}
 e^\mu \ 
   ={\displaystyle\frac{1-2m/r}{1+2\alpha m\cos\theta}}\;,\\[12pt]
 e^{-\lambda} 
   ={\displaystyle\Bigl(1-\frac{2m}{r}\Bigr)(1+2\alpha m \cos\theta)
   +\frac{m^2}{r^2}(1-\alpha^2 r^2)\sin^2\theta}\;.
 \end{array} 
  \label{mulambda1} 
  \end{equation}  
 Notice that the metric (\ref{brdiag-metric}) reduces to the Rindler metric (\ref{RindlMink}) when $\mu,\lambda\to0$. Moreover, since $\mu$ and $\lambda$ are independent of $\tau$ and $\ph$, it can be put explicitly into the boost-rotation symmetric form \cite{BicSch89b} by performing the transformation $\TBR=\pm\zBR\sinh\tau$, $\ZBR=\pm\zBR\cosh\tau$ in the static regions~II (and $\TBR=\pm\zBR\cosh\tau$, $\ZBR=\pm\zBR\sinh\tau$ in the regions of type~I). The presence of the boost symmetry is an important argument for the existence of acceleration in this space-time.

Of course the metric functions (\ref{mulambda1}) need to be re-written in terms of the variables $\zBR$ and $\rho$. For this purpose, it is convenient to introduce three auxiliary functions $R_i>0$, $i=1,2,3$ which will be given explicitly in (\ref{Ribrcoords}) below. In terms of these, the metric functions can be expressed in the forms 
\begin{equation} 
 e^\mu ={ R_1+R_2-2m \over R_1+R_2+2m} \,,\qquad 
 e^\lambda= { \Big((1-2\alpha m)R_1+(1+2\alpha m)R_2+4\alpha mR_3\Big)^2 \over
 4\,(1-4\alpha^2 m^2)^2\,R_1\,R_2}  \;. 
 \label{br-fns}
 \end{equation}

\section{Weyl and other coordinates in the static region}
\label{sc:tphconst}
     We have already introduced several sets of coordinates, specifically ${\{x,\,y\}}$ and ${\{r,\,\theta\}}$, which are simply related by (\ref{xytothetar}), and ${\{\zBR,\,\rho\}}$ given by (\ref{brtrans}). Each of these parametrize the surfaces 
${\tau,\varphi=\hbox{const.}}$ in the static regions~II. For the sake of completeness, we will also introduce yet another important set in this region, namely the Weyl coordinates ${\{\bar z,\bar\rho\}}$.

\begin{figure}[hpt] 
\centerline{ 
\parbox[t]{200pt}{\centering\small\includegraphics{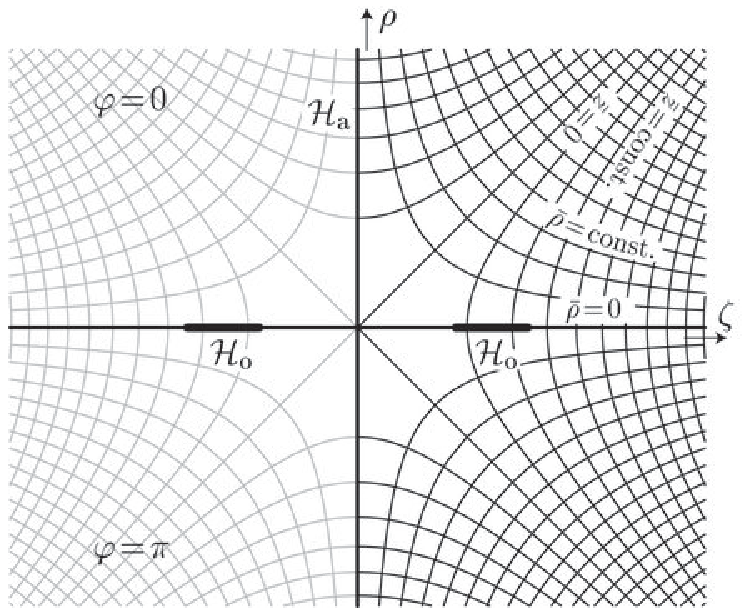}\\(a)}\quad
\parbox[t]{200pt}{\centering\small\includegraphics{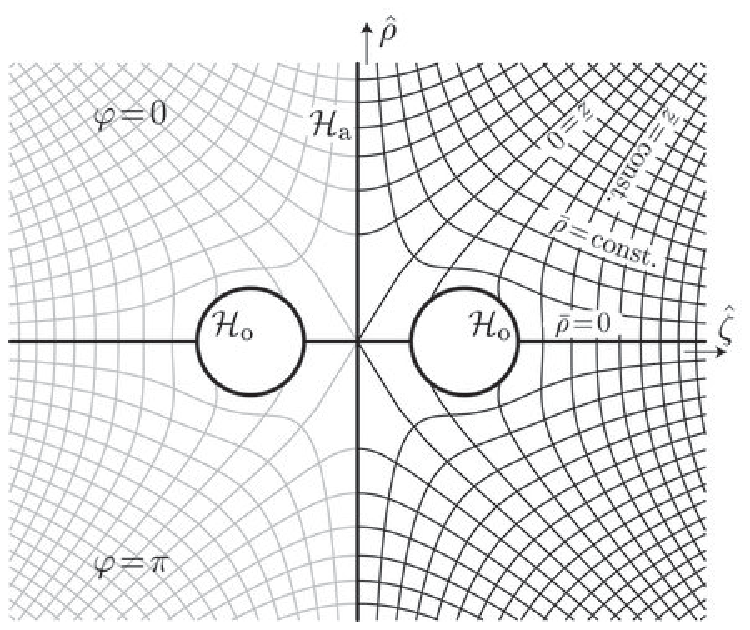}\\(b)}}
\caption{
 \label{fig:inBP}
 \small The coordinate lines of the Weyl coordinates ${\bar{z},\bar{\rho}}$ of (\ref{WeylBR}), plotted with respect to (a) boost-rotation symmetric coordinates $\{{\zBR,\rho\}}$ of (\ref{brtrans}), and (b) the grid ${\{\zBP, \rBP\}}$ given by the relations (\ref{BPcoor}). It can be seen that $\bar{\rho}$ does not penetrate the black hole horizons $\Ho$. The Weyl coordinates also cover only half of the section $\tau=\hbox{const.}$ Another set of the coordinates is drawn in grey on the left hand side of the diagram. }
\end{figure}

The relation between Weyl coordinates and the boost-rotation symmetric coordinates ${\zBR,\rho}$ is given by hyperbolic and parabolic orthogonal transformations, which are well-known from the flat 2-dimensional space \cite{BicSch89b}, namely 
 \begin{equation}
 \label{WeylBR}
  \left.\begin{array}{l}
  \bar z+\frac1{2\alpha} = \frac\alpha2\bigl(\zBR^2-\rho^2\bigr),\\[8pt] 
  \bar \rho = \alpha\,\zBR\rho\;,
  \end{array} \quad \right\} \quad 
  \left\{ \begin{array}{l}
  \sqrt\alpha\,\zBR=\sqrt{\textstyle\sqrt{{\bar\rho}^2
  +(\bar z+\frac1{2\alpha})^2}+(\bar z+\frac1{2\alpha})},\\
  \sqrt\alpha\,\rho=\sqrt{\textstyle\sqrt{{\bar\rho}^2+(\bar z+\frac1{2\alpha})^2}-(\bar z+\frac1{2\alpha})}.
  \end{array} \right. 
 \end{equation} 
 This is visualised in figure~\ref{fig:inBP}(a), with the corresponding picture for the similar grid (\ref{BPcoor}) given in \ref{fig:inBP}(b). These show the Weyl coordinate lines embedded in the same plane as that of figures \ref{fig:CM}(a) and \ref{fig:CM}(b).
The axis ${\bar\rho=0}$ of the Weyl coordinates covers both the axis of symmetry and the black hole and acceleration horizons. Indeed, different parts of the axis have the following meaning:
$$\begin{array}{lllc} 
 \hbox{Acceleration horizon}\ \Ha\ 
 & \leftrightarrow\quad 
 & \bar\rho=0, & \bar z\le-\frac1{2\alpha}, \\[3pt]
 \hbox{Inner axis} 
 &  \leftrightarrow\quad
 &  \bar\rho=0, & -\frac1{2\alpha}\le\bar z\le-m, \\[3pt]
 \hbox{Black hole horizon}\ \Ho
 & \leftrightarrow\quad 
 & \bar\rho=0, & -m\le\bar z<m, \\[3pt] 
 \hbox{Outer axis} 
  & \leftrightarrow\quad 
  & \bar\rho=0, & m\le\bar z.
 \end{array} $$

With (\ref{WeylBR}) and $\tau=\alpha t$, the metric (\ref{brdiag-metric}) takes the Weyl form 
 \begin{equation} 
 \d s^2 =  -e^{2U}\d t^2
 +e^{-2U+2\nu}\bigl(\d\bar\rho^2+\d\bar z^2\bigr) +e^{-2U}\bar\rho^2\d\varphi^2 ,
 \label{WeylMetric}
 \end{equation} 
 where $U$ and $\nu$ are functions of $\bar\rho$ and $\bar z$, given in terms of the auxiliary functions $R_i$ by 
 \begin{equation}
 \label{Unu-mulambda}
  \begin{array}{l}
 \qquad e^{2U}= \alpha\, 
 {\displaystyle {R_1+R_2-2m \over R_1+R_2+2m}}\,
 \Big(R_3 +(\bar z+\frac1{2\alpha})\Big), \\[8pt]
 e^{-2U+2\nu}= {\displaystyle {\Big(
 (1-2\alpha m)R_1+(1+2\alpha m)R_2+4\alpha mR_3\Big)^2 \over
 4\alpha\,(1-4\alpha^2 m^2)^2\,R_1\,R_2\,R_3}}.
 \end{array}
 \end{equation}

\begin{figure}[hpt]  
\bigskip    
\centerline{\includegraphics[scale=1.5]{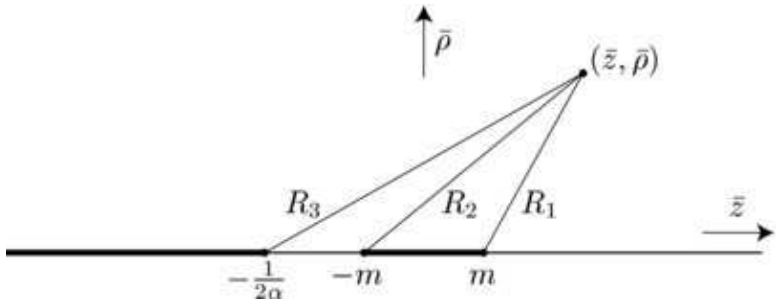}}
\caption{
 \label{fig:RinWeyl}
\small In Weyl coordinates, the axis $\bar\rho=0$ combines the acceleration horizon for ${\bar z<-1/2\alpha}$, the black hole horizon for ${\bar z\in(-m,m)}$, and the inner and outer parts of the axis of symmetry $\theta=0,\pi$. The auxiliary functions ${R_i}$ can be understood as distances from the edges of these horizons, measured in the sense of the unphysical Euclidian space in which Weyl coordinates ${\bar{z},\bar{\rho}}$ would be standard cylindrical coordinates.  }
\end{figure}

In this case, by introducing the points ${\bar z_1=m}$, ${\bar z_2=-m}$, ${\bar z_3=-1/2\alpha}$ on the Weyl axis $\bar\rho=0$, the auxiliary functions ${R_i>0}$, ${i=1,2,3}$ can be simply expressed respectively as 
 \begin{equation}
 \label{Ridef}
  R_i=\sqrt{(\bar z-\bar z_i)^2+\bar\rho^2}\;.
 \end{equation} 
 Thus, if we interpret the coordinates ${\{\bar\rho,\,\bar z\}}$ as cylindrical coordinates in an unphysical flat 2-space (following Bonnor \cite{Bonnor83}, Hong and Teo \cite{HongTeo03} and others), these functions can be understood as distances from the points ${\bar z=\bar z_i}$, $\bar\rho=0$, as illustrated in figure~\ref{fig:RinWeyl}. 
In fact, the functions $R_i$ can be expressed explicitly using (\ref{WeylBR}) and (\ref{brtrans}) in each of the following equivalent coordinate forms
 \begin{equation} 
  \begin{array}{l}
  {R_{1,2}}
   \displaystyle=\sqrt{( \bar z\mp m)^2+{\bar\rho}^2} \\[1.5ex]
  \qquad=\sqrt{\frac{1}{4\alpha^2}\Bigl(\alpha^2(\zBR^2+\rho^2)-(1\pm2\alpha m)\Bigr)^2+(1\pm2\alpha m)\rho^2} \\[1.5ex] 
   \displaystyle\qquad=\frac{r-m(1\mp\alpha r\mp\cos\theta-\alpha r\cos\theta)}{1+\alpha r\cos\theta} \;, \\[3ex]
  R_3\;\;
   =\sqrt{\Bigl(\bar z +\frac1{2\alpha}\Bigr)^2+{\bar\rho}^2} \\[1.5ex]
  \qquad=\frac12\,\alpha\,\bigl(\zBR^2+\rho^2)\  \\[1.5ex] 
   \displaystyle\qquad= \frac{(1-\alpha r\cos\theta)+2\alpha m(\cos\theta-\alpha r)}
   {2\alpha(1+\alpha r\cos\theta) \;.} 
  \end{array}
 \label{Ribrcoords} 
 \end{equation}

By substituting (\ref{brtrans}) into (\ref{WeylBR}), it is also found that the direct transformation between the original metric (\ref{BLCmetric}) and the Weyl form (\ref{WeylMetric}) is given by 
 \begin{equation}
 \bar z=-\frac{(\cos\theta\!+\!\alpha r)\Bigl(r\!-\!m(1\!-\!\alpha r\cos\theta)\Bigr)}
 {(1+\alpha r\cos\theta)^2}\;, \qquad
 \bar\rho =\frac{r\sin\theta\sqrt{P\,Q}}{(1+\alpha r\cos\theta)^2}\;. 
 \label{transf1}
 \end{equation} 
 Finally, in terms of the auxiliary functions (\ref{Ribrcoords}), the more complicated inverse transformation to (\ref{transf1}) can be written explicitly as 
  \begin{equation} 
  \begin{array}{l}
 {\displaystyle \quad\ \ r = {(1-2\alpha m)R_1 +(1+2\alpha m)R_2 
 +4\alpha mR_3 \over 
 (1-4\alpha^2m^2) -\alpha(1-2\alpha m)R_1 +\alpha(1+2\alpha m)R_2 
 +2\alpha R_3 } }, \\[3ex]
 \displaystyle \cos\theta = {(1-2\alpha m)(1+\alpha m)R_1 
 -(1+2\alpha m)(1-\alpha m)R_2 -4\alpha^2m^2R_3 \over 
 m\big[ (1-4\alpha^2m^2) -\alpha(1-2\alpha m)R_1 + \alpha(1+2\alpha m)R_2 
 +2\alpha R_3 \big] } \;,
  \end{array} 
  \label{invtransfxy} 
  \end{equation} 
 or equivalently: 
  $$ \cos\theta\,= \frac{(1-4\alpha^2m^2) +\alpha(1-2\alpha m)R_1 -\alpha(1+2\alpha m)R_2-2\alpha R_3} 
 {\alpha\big[ (1-2\alpha m)R_1 +(1+2\alpha m)R_2 +4\alpha m R_3\big]} \;. $$

\section{Conclusion}
     The $C$-metric has been presented above in a number of convenient coordinate systems. These have been based on the particularly useful form using bi-spherical coordinates given in (\ref{BLCmetric}) which makes explicit use of the simplified factorization structure that was introduced in~\cite{HongTeo03}. This also provides a form in which the physical properties and global structure of the space-time can more easily be studied and visualized.

In particular, the simple form of the metric (\ref{BLCmetric}) has enabled Kruskal--Szekeres type extensions to be explicitly obtained. To this, a conformal compactification has then been applied. This procedure has lead to the two-dimensional conformal diagrams given in figure~\ref{fig:CmtrCD} which describe the conformal structure of the manifold. Specifically, the complete space-time has been shown to represent an infinite sequence of alternating black holes and asymptotically flat regions.

Each asymptotically flat region is connected to a pair of causally separated black holes. The structure of these parts of the space-time has been illustrated in instructive three-dimensional causal pictures given in figures \ref{fig:3DCD}--\ref{fig:3Drcoor}, which have been constructed by appropriately combining the corresponding parts of the family of two-dimensional conformal diagrams of figure~\ref{fig:CmtrCD}. In particular, the characters of conformal infinity and the black hole and acceleration horizons have been clearly demonstrated.

Each pair of black holes appear to accelerate away from each other in opposite spatial directions along the axis of symmetry of the space-time. It is well known that this axis generally contains conical singularities which may be physically interpreted in terms of cosmic strings or struts whose tension or stress causes the acceleration of the black holes.

The metric (\ref{BLCmetric}) contains three arbitrary positive parameters, namely, $m$, $\alpha$ and $C$, which are explicitly related to the relevant physical quantities. Clearly, $m$ is related to the mass of the black holes in the sense that it is identical to the Schwarzschild mass parameter when $\alpha=0$. In addition, the space-time has a curvature singularity proportional to $m$ at $r=0$, which is located behind a black hole event horizon that is given exactly by $r=2m$ even when $\alpha\ne0$. In a similar way, $\alpha$ can be seen to be related to the acceleration of the black holes in the sense that it is exactly the acceleration of a corresponding test particle relative to a Minkowski background in the weak field limit in which ${m=0}$. Moreover, the space-time admits the Killing vector ${\partial_t}$ which is associated with a boost symmetry with an acceleration horizon at $r=1/\alpha$, even when $m\ne0$. The third, hidden, parameter~$C$ is defined only in terms of the range of~$\varphi$. It explicitly determines the distribution of conical singularities on the axis of symmetry. For any preferred distribution of strings and struts, $C$ is determined by the average of their deficit angles. Similarly, the dimensionless product $C\alpha m$ is given by the difference of the deficit angles, which can be interpret as the difference in the tensions of the strings (or struts) on opposite sides of the black holes.

The above interpretations of the parameters have partially been reached in particular limits. In the general case when ${m\neq0}$, it is difficult to measure the acceleration of the black holes since moving objects deform the space-time and drag local inertial frames. Also, the black hole is an extended object and its acceleration relative to a curved background cannot easily be defined. Similarly, when $\alpha\ne0$, it is difficult to uniquely determine the mass of each individual black hole since the space-time is not globally asymptotically flat. In fact, one cannot expect to distinguish effects due to acceleration from those due to gravitational fields. Nevertheless, the parameters $m$ and $\alpha$ employed above seem to be particularly appropriate for expressing the physical properties of the space-time and the various invariant geometrical quantities, such as those given in (\ref{BHarea})--(\ref{horsingtime}).

\section*{Acknowledgements}
    This work was supported in part by a grants from the EPSRC and by GACR 202/06/0041.


\begin{thebibliography}{99} 

\bibitem{EhlersKundt62} Ehlers, J. and Kundt, W. (1962). In {\sl Gravitation: an introduction to current research}, (ed. L. Witten). Wiley, New York. pp 49--101. 

\bibitem{Weyl17} Weyl, H. (1917). {\sl Ann. Physik}, {\bf 54}, 117. 

\bibitem{KinWal70} Kinnersley, W. and Walker, M. (1970). {\sl Phys. Rev. D} {\bf 2},
1359--1370. 

\bibitem{Bonnor83} Bonnor, W.B. (1983). {\sl Gen. Rel. Grav.} {\bf 15}, 535--551. 

\bibitem{FarZim80b} Farhoosh, H. and Zimmerman, R.L. (1980). 
{\sl Phys. Rev. D}, {\bf 21}, 2064--2074. 

\bibitem{Bicak85} Bi\v{c}\'ak, J. (1985).  In {\sl Galaxies, Axisymmetric
Systems and Relativity}, M.A.H. MacCallum, ed. (Cambridge University
Press, Cambridge), p 99. 

\bibitem{AshDra81} Ashtekar, A. and Dray. T. (1981). {\sl Commun. Math.
Phys.} {\bf 79}, 581--589. 

\bibitem{Pravdas00} Pravda, V. and Pravdov\'a, A. (2000). {\sl Czech. J. Phys.} {\bf 50}, 333--440. 

\bibitem{LetOli01} Letelier, P. S. and Oliveira, S. R. (2001). {\sl Phys. Rev.~D}, {\bf 64}, 064005, 1--9. 

\bibitem{HongTeo03} Hong, K. and Teo, E. (2003). {\sl Class. Quantum Grav.}, {\bf 20}, 3269--3277. 

\bibitem{GriPod05} Griffiths, J. B. and Podolsk\'y, J. (2005). {\sl Class. Quantum Grav.}, {\bf 22}, 3467--3479. 

\bibitem{GerHar82} Geroch, R, and Hartle, J. B. (1982). {\sl J. Math. Phys.}, {\bf 23}, 680--692. 

\bibitem{webpage} {\tt http://utf.mff.cuni.cz/\~{}krtous/physics/BlackHoles/}

\bibitem{BicSch89b} Bi\v{c}\'ak, J. and Schmidt, B. (1989$b$). {\sl Phys. Rev. D}, {\bf 40}, 1827--1853. 


\end{thebibliography}
\end{document}